\documentclass[pra,aps,epsfig,amsmath,amsbsy,showpacs]{revtex4}
\usepackage{amsmath,amsbsy,amsfonts}
\usepackage[dvips]{graphics,color}
\usepackage[dvips]{graphicx}
\begin{document}
\newcommand{\nn}{\nonumber \\}
\newcommand{\eps}{\ensuremath{\varepsilon}}
\newcommand{\me}{\mathrm{e}}
\newcommand{\im}{\ensuremath{\mathrm{i}}}
\newcommand{\bra}[1]{\langle #1 |}
\newcommand{\ket}[1]{| #1 \rangle}
\newcommand{\Bra}[1]{\left\langle #1 \left|}
\newcommand{\Ket}[1]{\right| #1 \right\rangle}
\newcommand{\bigbra}[1] {\big\langle #1\big|}
\newcommand{\bigket}[1] {\big|#1\big\rangle}
\newcommand{\Bigbra}[1] {\Big\langle #1\Big|}
\newcommand{\Bigket}[1] {\Big|#1\Big\rangle}
\newcommand{\dif}{\ensuremath{\mathrm{d}}}
\newcommand{\difbx}{\dif^3\bx}
\newcommand{\difbk}{\dif^3\bk}
\newcommand{\intD}[1]{\int\frac{\dif^3 #1}{(2\pi)^3}}
\newcommand{\intDD}[1]{\int\frac{\dif^4 #1}{(2\pi)^4}}
\newcommand{\intdinf}[1]{\int_{-\infty}^\infty\frac{\dif #1}{2\pi}}
\newcommand{\intbx}{\int\dif^3\bx}
\newcommand{\bx}{\boldsymbol{x}}
\newcommand{\br}{\boldsymbol{r}}
\newcommand{\bk}{\boldsymbol{k}}
\newcommand{\bp}{\boldsymbol{p}}
\newcommand{\bq}{\boldsymbol{q}}
\newcommand{\balpha}{\ensuremath{\boldsymbol{\alpha}}}
\newcommand{\dd}[1]{\frac{\partial}{\partial #1}}
\newcommand{\Partder}[1]{\frac{\partial }{\partial #1}}
\newcommand{\partder}[2]{\frac{\partial #1}{\partial #2}}
\newcommand{\Cov}{\mathrm{Cov}}
\newcommand{\eff}{_{\mathrm{eff}}}
\newcommand{\Red}{_{\mathrm{Red}}}
\newcommand{\SEp}{_{\mathrm{SEp}}}
\newcommand{\Irred}{\mathrm{Irred}}
\newcommand{\Sep}{\mathrm{Sep}}
\newcommand{\Nonsep}{\mathrm{Nonsep}}
\newcommand{\Irr}{_{\mathrm{Irr}}}
\newcommand{\RL}{_{\mathrm{L}}}
\newcommand{\QL}{_{\mathrm{QL}}}
\newcommand{\sgn}{\mathrm{sgn}}
\newcommand{\etc}{\mathrm{etc}}
\newcommand{\out}{\mathrm{out}}
\newcommand{\iin}{\mathrm{in}}
\newcommand{\inter}{\mathrm{int}}
\renewcommand{\rm}{\mathrm}
\newcommand{\mI}{\mathrm{I}}
\newcommand{\mS}{\mathrm{S}}
\newcommand{\mD}{\mathrm{D}}
\newcommand{\mG}{\mathrm{G}}
\newcommand{\mH}{\mathrm{H}}
\newcommand{\ext}{\mathrm{ext}}
\newcommand{\conn}{\mathrm{conn}}
\newcommand{\Counter}{\mathrm{Counter}}
\newcommand{\Linked}{\mathrm{linked}}
\newcommand{\linked}{\mathrm{linked}}
\newcommand{\mB}{\mathrm{B}}
\newcommand{\mC}{\mathrm{C}}
\newcommand{\SE}{\mathrm{SE}}
\newcommand{\G}{\mathrm{G}}
\newcommand{\R}{\mathrm{R}}
\newcommand{\sr}{\mathrm{SR}}
\newcommand{\bs}{\boldsymbol}
\newcommand{\bdot}{\boldsymbol\cdot}
\newcommand{\Q}{\mathcal{\boldsymbol{Q}}}
\newcommand{\OM}{\mathcal{\boldsymbol{\Om}}}
\renewcommand{\H}{V\eff}
\newcommand{\bH}{\bs{H}}
\newcommand{\U}{U}
\newcommand{\calH}{\mathcal{H}}
\newcommand{\dagg}{^{\dag}}
\newcommand{\intd}[1]{\int\frac{\dif #1}{2\pi}}
\newcommand{\half}{{\displaystyle\frac{1}{2}}}
\newcommand{\halfi}{{\displaystyle\frac{\im}{2}}}
\newcommand{\dint}{\int\!\!\!\int}
\newcommand{\ddint}{\dint\!\!\!\dint}
\newcommand{\dintd}[2]{\dint\frac{\dif #1}{2\pi}\,\frac{\dif #2}{2\pi}}
\newcommand{\ddintd}[4]{\ddint\frac{\dif #1}{2\pi}\,\frac{\dif #2}{2\pi}\,\frac{\dif #3}{2\pi}\,\frac{\dif #4}{2\pi}}
\newcommand{\ddt}{\frac{\partial}{\partial t}}
\newcommand{\intbr}{\int\dif\br\,}
\newcommand{\gamlim}{\ensuremath{\gamma\rightarrow 0}}
\newcommand{\vsp}{\vspace{0.5cm}}
\newcommand{\vvsp}{\vspace{0.25cm}}
\newcommand{\mvsp}{\vspace{-0.5cm}}
\newcommand{\mmvsp}{\vspace{-0.25cm}}
\newcommand{\mVsp}{\vspace{-1cm}}
\newcommand{\Vsp}{\vspace{1cm}}
\newcommand{\Wsp}{\vspace{2cm}}
\newcommand{\hsp}{\hspace{0.5cm}}
\newcommand{\Hsp}{\hspace{1cm}}
\newcommand{\HSP}{\hspace{2cm}}
\newcommand{\mmhsp}{\hspace{-0.25cm}}
\newcommand{\mhsp}{\hspace{-0.5cm}}
\newcommand{\mHsp}{\hspace{-1cm}}
\newcommand{\mHSP}{\hspace{-2cm}}
\newcommand{\ö}{\"{o}}
\newcommand{\Ö}{\"{O}}
\newcommand{\ä}{\"a}
\newcommand{\å}{\aa}
\newcommand{\Å}{\AA}
\newcommand{\hpsi}{\hat{\psi}}
\newcommand{\WU}{\widetilde{U}}
\renewcommand{\and}{\quad\mathrm{and}\quad}
\renewcommand{\it}{\textit}
\renewcommand{\bf}{\textbf}
\newcommand{\ul}{\underline}
\newcommand{\abs}[1]{|{#1}|}
\newcommand{\Abs}[1]{\big|{#1}\big|}
\newcommand{\eq}{\eqref}
\newcommand{\rarr}{\rightarrow}
\newcommand{\larr}{\leftarrow}
\newcommand{\lrarr}{\leftrightarrow}
\newcommand{\LRarr}{\Longleftrightarrow}
\newcommand{\Rarr}{\Rightarrow}
\newcommand{\Lrarr}{\Longrightarrow}
\newcommand{\rr}{r_{12}}
\newcommand{\brr}{\br_{12}}
\newcommand{\DF}{D_{\rm{F}\mu\nu}}
\newcommand{\SF}{S_{\rm{F}}}
\newcommand{\QED}{\rm{QED}}
\renewcommand{\sp}[2]{\bra{#1}{#2}\rangle}
\newcommand{\SP}[2]{\Bigbra{#1}{#2}\Big\rangle}
\newcommand{\qand}{\quad\rm{and}\quad}
\newcommand{\red}{\textcolor{red}}
\newcommand{\blue}{\textcolor{blue}}
\newcommand{\redbf}[1]{\bf{\red{#1}}}
\newcommand{\redbs}[1]{\bs{\red{#1}}}
\newcommand{\bluebf}[1]{\bf{\blue{#1}}}
\newcommand{\bluebs}[1]{\bs{\blue{#1}}}
\newcommand{\green}{\textcolor{green}}
\newcommand{\greenbf}[1]{\bf{\green{#1}}}
\newcommand{\ROmega}{\redbs{\Omega}}
\newcommand{\ROm}{\redbs{\Omega}}
\newcommand{\V}{\mathcal{V}}
\renewcommand{\v}{\nu}
\newcommand{\calK}{\mathcal{\kappa}}
\newcommand{\calKc}{I_c}
\newcommand{\calF}{\mathcal{F}}
\newcommand{\GQ}{\Gamma_Q}
\newcommand{\bGQ}{\bs{\Gamma_Q}}
\newcommand{\Gk}{\bs{\Gamma_Q}}
\newcommand{\Gv}{\Gamma_Q}
\newcommand{\TD}{T_\rm{D}}
\newcommand{\F}{\rm{F}}
\renewcommand{\P}{\bs{P}}
\newcommand{\Util}{\widetilde{U}}
\newcommand{\Om}{\Omega}
\newcommand{\Ombar}{\bar{\Omega}}
\newcommand{\VI}{V_{12}}
\newcommand{\VR}{V_{\rm{R}}}
\newcommand{\VRbar}{\bar{V}_{\rm{R}}}
\newcommand{\E}{{\mathcal{E}}}
\newcommand{\partdelta}[2]{\frac{\delta #1}{\delta #2}}
\newcommand{\pd}[1]{\frac{\delta #1}{\delta \E}}
\newcommand{\Partdelta}[1]{\frac{\delta }{\delta #1}}
\newcommand{\partdeltan}[3]{\frac{\delta^#1 #2}{\delta #3^#1}}
\newcommand{\pdn}[2]{\frac{\delta^#1 #2}{\delta\E^#1}}
 \newcommand{\ett}{^{(1)}}
 \newcommand{\limgam}{\lim_{\gamlim}}
 \newcommand{\Ugam}[1]{U_\gamma(#1,-\infty)}
\newcommand{\Ugamtil}[1]{\widetilde{U}_\gamma(#1,-\infty)}
\newcommand{\Ugamt}{\widetilde{U}_\gamma}
                                                  \newcommand{\tva}{^{(2)}}
                                                  \newcommand{\tre}{^{(3)}}
                                                  \newcommand{\fyr}{^{(4)}}
                                                  \newcommand{\enh}{^{(1/2)}}
                                                  \newcommand{\treh}{^{(3/2)}}
                                                  \newcommand{\femh}{^{(5/2)}}

\newcommand{\partdern}[3]{{\frac{\partial^#1 #2}
{\partial #3^#1}}}
  \newcommand{\MSC}{\rm{MSC}}
                                                  \newcommand{\Htil}{\widetilde{H}}
                                                  \newcommand{\Ubar}{\bar{U}}
                                                  \newcommand{\Hbar}{\bar{V}}
                                                  \newcommand{\Udot}{\dot{U}}
                                                  \newcommand{\Ubardot}{\dot{\Ubar}}
                                                  \newcommand{\Utildot}{\dot{\Util}}
                                                  \newcommand{\Cdot}{\dot{C}}
                                                  \newcommand{\Ombardot}{\dot{\Ombar}}
                                                  \newcommand{\bsdot}{\bs{\cdot}}

\newcommand{\npartdelta}[3]{\frac{\delta^#1 #2}{\delta #3^#1}}
\newcommand{\ip}[1]{| #1 \rangle \langle #1 |}
\newcommand{\con}{\mathrm{con}}
\newcommand{\ph}{\mathrm{ph}}


\newcommand{\LineH}[1]
{\linethickness{0.4mm}
\put(0,0){\line(1,0){#1}}
}
\newcommand{\LineS}[1]
{\linethickness{1mm}
\put(0,0){\line(1,0){#1}}
}

\newcommand{\LineWO}[1]
{\linethickness{0.5mm}
\put(0,0){\line(1,0){#1}}
}

\newcommand{\LineV}[1]
{\linethickness{0.4mm}
\put(0,0){\line(0,1){#1}}
}

\newcommand{\Linev}[1]
{\put(0.,0){\line(0,1){#1}}
}

\newcommand{\LineW}[1]
{\put(0.015,0){\line(0,1){#1}}
\put(0,0){\line(0,1){#1}}
\put(-0.015,0){\line(0,1){#1}}}

\newcommand{\LineHpt}[1]
{\put(0,0.015){\line(1,0){#1}}
\put(0,-0.015){\line(1,0){#1}}
\put(0,0){\circle*{0.15}}
\put(#1,0){\circle*{0.15}}}

\newcommand{\LineDl}[1]
{\put(0.012,-0.012){\line(-1,-1){#1}}\put(-0.012,0.012){\line(-1,-1){#1}}
\put(0.0,0.0){\line(-1,-1){#1}}}

\newcommand{\Linedl}[1]
{\put(0.01,0.01){\line(-1,-2){#1}}
\put(-0.01,-0.01){\line(-1,-2){#1}}}

\newcommand{\LineUr}[1]
{\put(0,0.015){\line(1,1){#1}}
\put(0,0){\line(1,1){#1}}
\put(0,-0.015){\line(1,1){#1}}}

\newcommand{\LineDr}[1]
{\put(0,0.01){\line(1,-1){#1}} \put(0,0){\line(1,-1){#1}}
\put(0,-0.01){\line(1,-1){#1}}}

\newcommand{\Linedr}[1]
{\put(0,0.01){\line(1,-3){#1}}
\put(0,0){\line(1,-3){#1}}
\put(0,-0.01){\line(1,-3){#1}}}

\newcommand{\Lineur}[1]
{\put(0,0.01){\line(1,3){#1}}
\put(0,0){\line(1,3){#1}}
\put(0,-0.01){\line(1,3){#1}}}

\newcommand{\DLine}[1]
{\put(0.,-0.05){\line(1,0){#1}}
\put(0.,0.05){\line(1,0){#1}}
\put(0,0){\circle*{0.1}}}

\newcommand{\Vector}[0]
{\thicklines\setlength{\unitlength}{1cm}\put(-0.13,0){\vector(-1,0){0}}}

\newcommand{\VectorR}[0]
{\thicklines\put(0.13,0.0){\vector(1,0){0}}}

\newcommand{\VectorUp}[0]
{\thicklines\setlength{\unitlength}{1cm}
\put(-0,0.12){\vector(0,0){0}}}

\newcommand{\VectorDn}[0]
{\thicklines\setlength{\unitlength}{1cm}
\put(0.012,-0.12){\vector(0,-1){0}}
\put(-0.012,-0.12){\vector(0,-1){0}}}

\newcommand{\VectorDl}[0]
{\thicklines
\setlength{\unitlength}{1cm}
\put(-0.092,-0.076){\vector(-1,-1){0}}
\put(-0.076,-0.092){\vector(-1,-1){0}}
}

\newcommand{\VectorDr}[0]
{\thicklines
\setlength{\unitlength}{1cm}
\put(0.076,-0.092){\vector(1,-1){0}}
\put(0.092,-0.076){\vector(1,-1){0}}
}

\newcommand{\Vectordr}[0]
{\setlength{\unitlength}{1cm}
\put(0.022,0.112){\vector(1,-3){0}}
\put(0.052,0.118){\vector(1,-3){0}}
}

\newcommand{\Vectorur}[0]
{\setlength{\unitlength}{1cm}
\put(0.04,-0.062){\vector(1,3){0}}
\put(0.06,-0.068){\vector(1,3){0}}
}

\newcommand{\VectorUr}[0]
{\put(0.195,0.705){\vector(1,1){0}}
 \put(0.22,0.76){\vector(1,1){0}}
\put(0.155,0.735){\vector(1,1){0}} }

\newcommand{\VectorUl}[0]
{\put(-0.23,-0.02){\vector(-1,1){0}}
\put(-0.19,-0.03){\vector(-1,1){0}}
\put(-0.22,-0.06){\vector(-1,1){0}}
}

\newcommand{\Wector}[0]
{\put(-0.15,0)\Vector\put(0.15,0)\Vector}

\newcommand{\WectorUp}[0]
{\put(0,0.125)\VectorUp\put(0,-0.125)\VectorUp}

\newcommand{\WectorDn}[0]
{\put(0,0.125)\VectorDn\put(0,-0.125)\VectorDn}

\newcommand{\WectorDl}[0]
{\put(0.1,0.1)\VectorDl\put(-0.1,-0.1)\VectorDl}

\newcommand{\EllineH}[4]
{\put(0,0){\LineH{#1}}
\put(#2,0){\Vector}
\put(#2,0.45){\makebox(0,0){$#3$}}
\put(#2,-0.35){\makebox(0,0){$#4$}}}

\newcommand{\lline}[4]
{\put(0,0){\LineV{#1}} \put(-0.3,#2){\makebox(0,0){$#3$}}
\put(0.4,#2){\makebox(0,0){$#4$}}}

\newcommand{\Elline}[4]
{\put(0,0){\LineV{#1}} \put(0,#2){\VectorUp}
\put(-0.3,#2){\makebox(0,0){$#3$}}
\put(0.4,#2){\makebox(0,0){$#4$}}}

\newcommand{\DElline}[4]
{\put(0,0){\LineV{#1}}
\put(0,#2){\WectorUp}
\put(-0.3,#2){\makebox(0,0){$#3$}}
\put(0.3,#2){\makebox(0,0){$#4$}}}

\newcommand{\DEllineDn}[4]
{\put(0,0){\LineV{#1}}
\put(0,#2){\WectorDn}
\put(-0.25,#2){\makebox(0,0){$#3$}}
\put(0.25,#2){\makebox(0,0){$#4$}}}

\newcommand{\Ellinet}[4]
{\put(0,0){\Linev{#1}} \put(0,#2){\vector(0,1){0}}
\put(-0.35,#2){\makebox(0,0){$#3$}}
\put(0.35,#2){\makebox(0,0){$#4$}}}
\newcommand{\EllineT}[4]
{\put(0,0){\LineW{#1}}
\put(0,#2){\VectorUp}
\put(-0.35,#2){\makebox(0,0){$#3$}}
\put(0.35,#2){\makebox(0,0){$#4$}}}

\newcommand{\EllineDnt}[4]
{\put(0,0){\Linev{#1}}
\put(0,#2){\VectorDn}
\put(-0.35,#2){\makebox(0,0){$#3$}}
\put(0.35,#2){\makebox(0,0){$#4$}}}

\newcommand{\EllineDn}[4]
{\put(0,0){\LineV{#1}} \put(0,#2){\VectorDn}
\put(-0.35,#2){\makebox(0,0){$#3$}}
\put(0.35,#2){\makebox(0,0){$#4$}}}

\newcommand{\EllineDl}[4]
{\put(0,0){\LineDl{#1}} \put(-#2,-#2){\VectorDl}
\put(-0.2,0.4){\makebox(-#1,-#1){$#3$}}
\put(0.2,-0.2){\makebox(-#1,-#1){$#4$}}}

\newcommand{\Ellinedl}[4]
{\put(0.01,0.01){\line(-1,-2){#1}}
\put(-0.01,-0.01){\line(-1,-2){#1}}
\thicklines\put(-0.05,-0.1){\vector(-1,-2){#2}}
\put(-0.2,0.2){\makebox(-#1,-#1){$#3$}}
\put(0.2,-0.2){\makebox(-#1,-#1){$#4$}}}

\newcommand{\Ellinedr}[4]
{\put(0.01,0.01){\line(1,-2){#1}}
\put(-0.01,-0.01){\line(1,-2){#1}}
\thicklines\put(0,0){\vector(1,-2){#2}}
\put(0.2,-0.3){\makebox(#1,-#1){$#3$}}
\put(0,-1){\makebox(#1,-#1){$#4$}}}

\newcommand{\EllineA}[7]
{\put(0.0,0.0){\line(#1,#2){#3}}
\put(0.005,0.0){\line(#1,#2){#3}}
\put(-0.005,0.0){\line(#1,#2){#3}}
\put(0,0){\vector(#1,#2){#4}}
\put(0.010,0){\vector(#1,#2){#4}}
\put(-0.010,0){\vector(#1,#2){#4}}
\put(#6,#7){\makebox(0,0){$#5$}}}

\newcommand{\EllineDr}[4]
{\put(0,0){\LineDr{#1}} \put(#2,-#2){\VectorDr}
\put(#2,-#2){\makebox(-0.5,-0.5){$#3$}}
\put(#2,-#2){\makebox(0.5,0.5){$#4$}}}

\newcommand{\EllinedR}[5]
{\put(0,0){\line(1,-3){#1}}
\put(0.014,0){\line(1,-3){#1}}
\put(-0.014,0){\line(1,-3){#1}}
\put(#2,-#3){\makebox(0,0){{\Vectordr}}}
\put(#2,-#3){\makebox(-0.5,-0.5){$#4$}}
\put(#2,-#3){\makebox(0.5,0.5){$#5$}}}

\newcommand{\EllineuR}[5]
{\put(0,0){\line(1,3){#1}}
\put(0.014,0){\line(1,3){#1}}
\put(-0.014,0){\line(1,3){#1}}
\put(#2,#3){\makebox(0,0){\Vectorur}}
\put(#2,#3){\makebox(-0.5,-0.5){$#4$}}
\put(#2,#3){\makebox(0.5,0.5){$#5$}}}


\newcommand{\Ellineur}[5]
{\put(0,0){\Lineur{#1}}
\put(#2,#3){\makebox(0,0){\Vectorur}}
\put(#2,#3){\makebox(-0.5,0){$#4$}}
\put(#2,#3){\makebox(0.5,0){$#5$}}}

\newcommand{\EllineUr}[4]
{\put(0,0){\LineUr{#1}}
\put(#2,#2){\makebox(-0.35,-0.35){\VectorUr}}
\put(-0.2,0.4){\makebox(#1,#1){$#3$}}
\put(0.2,-0.3){\makebox(#1,#1){$#4$}}}

\newcommand{\DEllineDl}[4]
{\put(0,0){\LineDl{#1}}
\put(-#2,-#2){\WectorDl}
\put(-0.25,0.25){\makebox(-#1,-#1){$#3$}}
\put(0.25,-0.25){\makebox(-#1,-#1){$#4$}}}

\newcommand{\Ebox}[2]
{\put(0,0){\LineH{#1}}
\put(0,#2){\LineH{#1}}
\put(0,0){\LineV{#2}}
\put(#1,0){\LineV{#2}}}

\newcommand{\dashH}
{\multiput(0.05,0)(0.25,0){5}{\line(1,0){0.15}}}

\newcommand{\dash}[1]
{\multiput(0.05,0)(0.25,0){#1}{\line(1,0){0.15}}}

\newcommand{\dashV}[1]
{\multiput(0.05,0)(0,0.25){#1}{\line(0,1){0.15}}}

\newcommand{\dashHp}
{\multiput(0.05,0)(0.25,0){6}{\line(1,0){0.15}}}

\newcommand{\DashH}
{\multiput(0.05,0)(0.25,0){10}{\line(1,0){0.15}}}

\newcommand{\dashHnum}[2]
{\multiput(0.05,0)(0.25,0){5}{\line(1,0){0.15}}
\put(-0.25,0){\makebox(0,0){$#1$}}
\put(1.5,0){\makebox(0,0){$#2$}}}

\newcommand{\dashHnuma}[2]
{\multiput(0.05,0)(0.25,0){5}{\line(1,0){0.15}}
\put(0.25,0.25){\makebox(0,0){$#1$}}
\put(1,0.25){\makebox(0,0){$#2$}}}

\newcommand{\dashHnumu}[2]
{\multiput(0.05,0)(0.25,0){5}{\line(1,0){0.15}}
\put(0.25,-0.25){\makebox(0,0){$#1$}}
\put(1,-0.25){\makebox(0,0){$#2$}}}

\newcommand{\Potint}
{\put(0,0)\dashH \put(1.35,0){\makebox(0,0){x}}
\put(0,0){\circle*{0.15}}}

\newcommand{\potint}
{\multiput(0.05,0)(0.25,0){3}{\line(1,0){0.15}}
\put(0.85,0){\makebox(0,0){x}}
\put(0,0){\circle*{0.15}}}

\newcommand{\PotintS}
{\put(0,0){\dash{4}} \put(1,0){\makebox(0,0){$\times$}}
\put(0,0){\circle*{0.1}}}

\newcommand{\PotintL}
{\put(-1.25,0)\dashH
\put(-1.35,0){\makebox(0,0){x}}
\put(0,0){\circle*{0.15}}}

\newcommand{\Effpot}
{\put(0,0)\dashH
\put(1.35,0){\makebox(0,0){x}}
\put(1.35,0){\circle{0.3}}
\put(0,0){\circle*{0.15}}}

\newcommand{\effpot}
{\multiput(0.05,0)(0.25,0){3}{\line(1,0){0.15}}
\put(0.85,0){\makebox(0,0){x}}
\put(0.85,0){\circle{0.3}}
\put(0,0){\circle*{0.1}}}

\newcommand{\EffpotL}
{\put(-1.25,0)\dashH
\put(0,0){\makebox(0,0){x}}
\put(-1.25,0){\circle{0.3}}
\put(-1.25,0){\circle*{0.15}}}

\newcommand{\Triang}
{\put(0,0){\line(2,1){0.5}}
\put(0,0){\line(2,-1){0.5}}
\put(0.5,-0.25){\line(0,1){0.5}}}

\newcommand{\TriangL}
{\put(0,0){\line(-2,1){0.5}}
\put(0,0){\line(-2,-1){0.5}}
\put(-0.5,-0.25){\line(0,1){0.5}}}

\newcommand{\hfint}
{\put(0,0)\dashH
\put(1.25,0){\makebox(0,0){\Triang}}
\put(0,0){\circle*{0.15}}}

\newcommand{\hfintL}
{\put(-1.25,0)\dashH
\put(-1.25,0){\makebox(0,0){\TriangL}}
\put(0,0){\circle*{0.15}}}

\newcommand{\VPloop}[1]
{\put(0,0){\circle{1}}
\put(0,0.0){\circle{1.}}
\put(0.02,0){\circle{1.}}
\put(0,0.02){\circle{1.}}
\put(0,-0.02){\circle{1}}
\put(-0.02,0){\circle{1}}
\put(0.52,0.05){\VectorDn}
\put(0.85,0){\makebox(0,0){$#1$}}}

\newcommand{\VPloopt}[1]
{\put(0,0){\circle{1}}
\put(0.52,0.05){\VectorDn}
\put(0.75,0){\makebox(0,0){$#1$}}}

\newcommand{\VPloopL}[1]
{\put(0,0){\circle{1}}
\put(0.01,0.){\circle{1}}\put(-0.01,0){\circle{1}}
\put(0,0.01){\circle{1}}\put(0,-0.01){\circle{1}}
\put(-0.5,0.05){\VectorDn}
\put(-0.75,0){\makebox(0,0){$#1$}}}

\newcommand{\VPloopLt}[1]
{\put(0,0){\circle{1}}
\put(-0.5,0){\VectorDn}
\put(-0.75,0){\makebox(0,0){$#1$}}}

\newcommand{\VPloopLR}[2]
{\put(0,0){\circle{1}}\put(-0.01,0){\circle{1}}
\put(0,0.01){\circle{1}}\put(0,-0.01){\circle{1}}
\put(-0.5,0){\VectorDn}
\put(0.5,0){\VectorUp}
\put(-0.75,0){\makebox(0,0){$#1$}}
\put(0.75,0){\makebox(0,0){$#2$}}}

\newcommand{\VPloopLRt}[2]
{\put(0,0){\circle{1}}
\put(-0.5,0){\VectorDn}
\put(0.5,0){\VectorUp}
\put(-0.75,0){\makebox(0,0){$#1$}}
\put(0.75,0){\makebox(0,0){$#2$}}}

\newcommand{\VPloopD}[2]
{\put(0,0){\circle{1}}
\put(0.01,0.){\circle{1}}\put(-0.01,0){\circle{1}}
\put(0.,0.){\circle{1}}\put(-0.01,0){\circle{1}}
\put(0,0.01){\circle{1}}\put(0,-0.01){\circle{1}}
\put(0,0.52){\VectorR}
\put(0,-0.52){\Vector}
\put(0,0.8){\makebox(0,0){$#1$}}
\put(0,-0.8){\makebox(0,0){$#2$}}}

\newcommand{\VPloopDt}[2]
{\put(0,0){\circle{1}}
\put(0,0.5){\VectorR}
\put(0.05,-0.47){\Vector}
\put(0,0.8){\makebox(0,0){$#1$}}
\put(0,-0.8){\makebox(0,0){$#2$}}}

\newcommand{\Loop}[2]
{\put(0,0){\oval(0.6,1.25)}\put(0.01,0.01){\oval(0.6,1.25)}\put(-0.01,-0.01){\oval(0.6,1.25)}
\put(0.3,0){\VectorUp}
\put(-0.3,0){\VectorDn}
\put(-0.65,0){\makebox(0,0){$#1$}}
\put(0.65,0){\makebox(0,0){$#2$}}}

\newcommand{\HFexch}[1]
{\put(0,0)\dashH
\qbezier(0,0.01)(0.625,0.515)(1.25,0.015)
\qbezier(0,-0.01)(0.625,0.485)(1.25,-0.015)
\put(0.625,0.26){\Vector}
\put(0.625,0.5){\makebox(0,0){$#1$}}
\put(0,0){\circle*{0.15}}
\put(1.25,0){\circle*{0.15}}}

\newcommand{\HFexcht}[1]
{\put(0,0)\dashH
\qbezier(0,0.01)(0.625,0.515)(1.25,0.015)
\put(0.625,0.26){\Vector}
\put(0.625,0.5){\makebox(0,0){$#1$}}
\put(0,0){\circle*{0.15}}
\put(1.25,0){\circle*{0.15}}}

\newcommand{\Dashpt}[1]
{\multiput(0,-0.6)(0,0.25){13}{\line(0,1){0.15}}
\put(0,0){\circle*{0.15}}
\put(0,#1){\circle*{0.15}}}

\newcommand{\photonH}[3]
{\qbezier(0,0)(0.08333,0.125)(0.1666667,0)
\qbezier(0.1666667,0)(0.25,-0.125)(0.3333333,0)
\qbezier(0.3333333,0)(0.416667,0.125)(0.5,0)
\qbezier(0.5,0)(0.583333,-0.125)(0.666667,0)
\qbezier(0.666667,0)(0.75,0.125)(0.833333,0)
\qbezier(0.833333,0)(0.916667,-0.125)(1,0)
\qbezier(1,0)(1.083333,0.125)(1.166667,0)
\qbezier(1.166667,0)(1.25,-0.125)(1.333333,0)
\qbezier(1.333333,0)(1.416667,0.125)(1.5,0)
\put(0.75,0.){\VectorR} \put(0.75,0.35){\makebox(0,0){$#1$}}
\put(0,0){\circle*{0.1}} \put(1.5,0){\circle*{0.1}}
\put(-0.5,0){\makebox(0,0){#2}} \put(2,0){\makebox(0,0){#3}}}

\newcommand{\photon}[3]
{\qbezier(0,0)(0.08333,0.125)(0.1666667,0)
\qbezier(0.1666667,0)(0.25,-0.125)(0.3333333,0)
\qbezier(0.3333333,0)(0.416667,0.125)(0.5,0)

\qbezier(0.5,0)(0.583333,-0.125)(0.666667,0)
\qbezier(0.666667,0)(0.75,0.125)(0.833333,0)
\qbezier(0.833333,0)(0.916667,-0.125)(1,0)

\qbezier(1,0)(1.083333,0.125)(1.166667,0)
\qbezier(1.166667,0)(1.25,-0.125)(1.333333,0)
\qbezier(1.333333,0)(1.416667,0.125)(1.5,0)

\qbezier(1.5,0)(1.583333,-0.125)(1.666667,0)
\qbezier(1.666667,0)(1.75,0.125)(1.833333,0)
\qbezier(1.833333,0)(1.916667,-0.125)(2,0)
\put(1,0.0){\VectorR}
\put(1,0.35){\makebox(0,0){$#1$}}
\put(0,0){\circle*{0.15}}
\put(2,0){\circle*{0.15}}
\put(-0.35,0){\makebox(0,0){#2}}
\put(2.35,0){\makebox(0,0){#3}}}

\newcommand{\Photon}[3]
{\put(0,0){\photon{}{}{}}\put(0.05,0.05){\photon{}{}{}}
\put(-0.05,-0.05){\photon{}{}{}}}

\newcommand{\photonHS}[4]
{\qbezier(0,0)(0.08333,0.125)(0.1666667,0)
\qbezier(0.1666667,0)(0.25,-0.125)(0.3333333,0)
\qbezier(0.3333333,0)(0.416667,0.125)(0.5,0)
\qbezier(0.5,0)(0.583333,-0.125)(0.666667,0)
\qbezier(0.666667,0)(0.75,0.125)(0.833333,0)
\qbezier(0.833333,0)(0.916667,-0.125)(1,0)
\put(0.5,0.025){\VectorR}
\put(0.5,0.35){\makebox(0,0){$#1$}}
\put(0,0){\circle*{0.15}}
\put(1,0){\circle*{0.15}}
\put(0,-0.5){\makebox(0,0){#2}}
\put(1,-0.5){\makebox(0,0){#3}}}

\newcommand{\photonNE}[3]
{\qbezier(0,0)(0.22,-0.02)(0.2,0.2)
\qbezier(0.2,0.2)(0.18,0.42)(0.4,0.4)
\qbezier(0.4,0.4)(0.62,0.38)(0.6,0.6)
\qbezier(0.6,0.6)(0.58,0.82)(0.8,0.8)
\qbezier(0.8,0.8)(1.02,0.78)(1,1)
\qbezier(1,1)(0.98,1.22)(1.2,1.2)
\qbezier(1.2,1.2)(1.42,1.18)(1.4,1.4)
\qbezier(1.4,1.4)(1.38,1.62)(1.6,1.6)
\qbezier(1.6,1.6)(1.82,1.58)(1.8,1.8)
\qbezier(1.8,1.8)(1.78,2.02)(2,2)
\put(1,1){\makebox(0.05,-0.2){\VectorUp}} \put(0,0){\circle*{0.1}}
\put(2,2){\circle*{0.1}} \put(1,1){\makebox(-0.6,0.4){$#1$}}
\put(-0.35,-1){\makebox(0,2){$#2$}}
\put(2.35,1){\makebox(0,2){$#3$}}}

\newcommand{\photonNNE}[3]
{\qbezier(0,0)   (0.28,-0.02)(0.2,0.3)
\qbezier(0.2,0.3)(0.12,0.52)(0.4,0.6)
\qbezier(0.4,0.6)(0.68,0.58)(0.6,0.9)
\qbezier(0.6,0.9)(0.52,1.12)(0.8,1.2)
\qbezier(0.8,1.2)(1.08,1.18)(1,1.5) \qbezier(1,1.5)
(0.92,1.72)(1.2,1.8) \qbezier(1.2,1.8)(1.48,1.86)(1.4,2.1)
\qbezier(1.4,2.1)(1.365,2.24)(1.6,2.4)
\qbezier(1.6,2.4)(1.835,2.46)(1.8,2.7)
\qbezier(1.8,2.7)(1.765,2.84)(2,3)
\put(0.6,0.8){\makebox(0,0){\VectorUp}} \put(0,0){\circle*{0.1}}
\put(2,3){\circle*{0.1}} \put(1,0.8){\makebox(0,0){$#1$}}
\put(-0.35,0){\makebox(0,0){$#2$}}
\put(2.35,3){\makebox(0,0){$#3$}}}

\newcommand{\photonENE}[3]
{\qbezier(0,0)(0.17,-0.04)(0.2,0.1)
\qbezier(0.2,0.1)(0.23,0.32)(0.4,0.2)
\qbezier(0.4,0.2)(0.57,0.16)(0.6,0.3)
\qbezier(0.6,0.3)(0.63,0.52)(0.8,0.4)
\qbezier(0.8,0.4)(0.97,0.36)(1,0.5)
\qbezier(1,0.5)(1.03,0.72)(1.2,0.6)
\qbezier(1.2,0.6)(1.37,0.56)(1.4,0.7)
\qbezier(1.4,0.7)(1.43,0.92)(1.6,0.8)
\qbezier(1.6,0.8)(1.77,0.76)(1.8,0.9)
\qbezier(1.8,0.9)(1.83,1.12)(2,1)
\put(1.25,0.85){\makebox(-0.1,0.06){\VectorR}}
\put(1,0.85){\makebox(0,-0){$#1$}}
\put(-0.35,-1){\makebox(0,2){$#2$}}
\put(2.35,0){\makebox(0,2){$#3$}}}

\newcommand{\photonNW}[3]
{\qbezier(0,0)(-0.22,-0.02)(-0.2,0.2)
\qbezier(-0.2,0.2)(-0.18,0.42)(-0.4,0.4)
\qbezier(-0.4,0.4)(-0.62,0.38)(-0.6,0.6)
\qbezier(-0.6,0.6)(-0.58,0.82)(-0.8,0.8)
\qbezier(-0.8,0.8)(-1.02,0.78)(-1,1) \qbezier(-1,1)
(-0.98,1.22)(-1.2,1.2) \qbezier(-1.2,1.2)(-1.42,1.18)(-1.4,1.4)
\qbezier(-1.4,1.4)(-1.38,1.62)(-1.6,1.6)
\qbezier(-1.6,1.6)(-1.82,1.58)(-1.8,1.8)
\qbezier(-1.8,1.8)(-1.78,2.02)(-2,2)
\put(-1,1){\makebox(0,-0.2){\VectorUp}} \put(0,0){\circle*{0.1}}
\put(-2,2){\circle*{0.1}} \put(-1,1){\makebox(0.4,0.7){$#1$}}
\put(-2.35,2){\makebox(0,0){$#2$}}
\put(0.35,0){\makebox(0,0){$#3$}}}

\newcommand{\photonSEst}[3]
{\qbezier(0,0)(-0.22,-0.02)(-0.2,0.2)
\qbezier(-0.2,0.2)(-0.18,0.42)(-0.4,0.4)
\qbezier(-0.4,0.4)(-0.62,0.38)(-0.6,0.6)
\qbezier(-0.6,0.6)(-0.58,0.82)(-0.8,0.8)
\qbezier(-0.8,0.8)(-1.02,0.78)(-1,1) \qbezier(-1,1)
(-0.98,1.22)(-1.2,1.2) \qbezier(-1.2,1.2)(-1.42,1.18)(-1.4,1.4)
\qbezier(-1.4,1.4)(-1.38,1.62)(-1.6,1.6)
\qbezier(-1.6,1.6)(-1.82,1.58)(-1.8,1.8)
\qbezier(-1.8,1.8)(-1.78,2.02)(-2,2)
\put(-1,1){\makebox(0,0){\VectorDn}} \put(0,0){\circle*{0.1}}
\put(-2,2){\circle*{0.1}} \put(-1,1){\makebox(0.4,0.7){$#1$}}
\put(-2.35,2){\makebox(0,0){$#2$}}
\put(0.35,0){\makebox(0,0){$#3$}}}

\newcommand{\photonWNW}[3]
{\qbezier(0,0)(-0.17,-0.04)(-0.2,0.1)
\qbezier(-0.2,0.1)(-0.23,0.32)(-0.4,0.2)
\qbezier(-0.4,0.2)(-0.57,0.16)(-0.6,0.3)
\qbezier(-0.6,0.3)(-0.63,0.52)(-0.8,0.4)
\qbezier(-0.8,0.4)(-0.97,0.36)(-1,0.5)
\qbezier(-1,0.5)(-1.03,0.72)(-1.2,0.6)
\qbezier(-1.2,0.6)(-1.37,0.56)(-1.4,0.7)
\qbezier(-1.4,0.7)(-1.43,0.92)(-1.6,0.8)
\qbezier(-1.6,0.8)(-1.77,0.76)(-1.8,0.9)
\qbezier(-1.8,0.9)(-1.83,1.12)(-2,1)
\put(-1,1){\makebox(-0.2,-0.4){$\;$\VectorR}}
\put(0,0){\circle*{0.1}} \put(-2,1){\circle*{0.1}}
\put(-1,0.5){\makebox(0.6,0.4){$#1$}}
\put(0.35,-1){\makebox(0,2){$#3$}}
\put(-2.35,0){\makebox(0,2){$#2$}}}

\newcommand{\Crossphotons}[6]
{\put(0,0){\photonNE{#5}{}{}}
\put(2,0){\photonNW{#6}{}{}}
\put(-0.35,0){\makebox(0,0){$#1$}}
\put(2.35,2){\makebox(0,0){$#2$}}
\put(2.35,0){\makebox(0,0){$#3$}}
\put(-0.35,2){\makebox(0,0){$#4$}}
}

\newcommand{\photonNe}[3]
{\qbezier(0,0)(0.22,-0.02)(0.2,0.2)
\qbezier(0.2,0.2)(0.18,0.42)(0.4,0.4)
\qbezier(0.4,0.4)(0.62,0.38)(0.6,0.6)
\qbezier(0.6,0.6)(0.58,0.82)(0.8,0.8)
\qbezier(0.8,0.8)(1.02,0.78)(1,1)
\qbezier(1,1)(0.98,1.22)(1.2,1.2)
\put(0.75,0.75){\makebox(-0.6,0.4){$#1$}}
\put(-0.35,0){\makebox(0,0){$#2$}}
\put(1.85,1.5){\makebox(0,0){$#3$}}}

\newcommand{\PhotonNe}[3]
{\put(0,0){\photonNe{}{}{}}\put(0.05,-0.05){\photonNe{}{}{}}
\put(-0.05,0.05){\photonNe{}{}{}}
\put(0.75,0.75){\makebox(-0.6,0.4){$#1$}}
\put(-0.35,0){\makebox(0,0){$#2$}}
\put(1.85,1.5){\makebox(0,0){$#3$}}}

\newcommand{\photonNw}[3]
{\qbezier(0,0)(0.02,0.22)(-0.2,0.2)
\qbezier(-0.2,0.2)(-0.42,0.18)(-0.4,0.4)
\qbezier(-0.4,0.4)(-0.38,0.62)(-0.6,0.6)
\qbezier(-0.6,0.6)(-0.82,0.58)(-0.8,0.8)
\qbezier(-0.8,0.8)(-0.78,1.02)(-1,1) \qbezier(-1,1)
(-1.22,0.98)(-1.2,1.2) \qbezier(-1.2,1.2)(-1.18,1.42)(-1.4,1.4)
\qbezier(-1.4,1.4)(-1.5,1.38)(-1.5,1.5)
\put(-0.5,0.8){\makebox(0,-0){\VectorR}}
\put(-0.7,0.7){\makebox(1,0.3){$#1$}}
\put(0.35,0){\makebox(0,0){$#3$}}
\put(-1.85,1.5){\makebox(0,0){$#2$}} }

\newcommand{\elstat}[3]
{\multiput(0.06,0)(0.25,0){8}{\line(1,0){0.15}}
\put(1,0.35){\makebox(0,0){$#1$}} 
\put(-0.35,0){\makebox(0,0){#2}}
\put(2.35,0){\makebox(0,0){#3}}}

\newcommand{\Melstat}[1]
{\multiput(0,0)(0,0.2){#1}{\elstat{}{}{}}}

\newcommand{\Multiline}[3]
{\linethickness{0.2mm} \put(0,-0.1){\line(1,0){#1}}
 \put(0,0){\line(1,0){#1}}\put(0,0.1){\line(1,0){#1}}
 \put(-0.35,0){\makebox(0,0){#2}}
\put(2.35,0){\makebox(0,0){#3}}}

\newcommand{\elstatH}[3]
{\multiput(0.06,0)(0.25,0){6}{\line(1,0){0.15}}
\put(0.75,0.25){\makebox(0,0){$#1$}} 
\put(-0.35,0){\makebox(0,0){#2}}
\put(2.35,0){\makebox(0,0){#3}}}

\newcommand{\BreitH}[3]
{\multiput(0.15,0)(0.3,0){5}{\circle*{0.1}}
\put(0.75,0.25){\makebox(0,0){$#1$}} 
\put(-0.35,0){\makebox(0,0){#2}}
\put(2.35,0){\makebox(0,0){#3}}}

\newcommand{\RetBreitDH}[3]
{\small\multiput(0.15,0.3)(0.3,-0.15){5}{\circle*{0.1}}
\put(0.75,0.25){\makebox(0,0){$#1$}}
\put(-0.35,0){\makebox(0,0){#2}} \put(2.35,0){\makebox(0,0){#3}}}

\newcommand{\RetBreitH}[3]
{\small\multiput(0.15,-0.3)(0.3,0.15){5}{\circle*{0.1}}
\put(0.75,0.25){\makebox(0,0){$#1$}}
\put(-0.35,0){\makebox(0,0){#2}} \put(2.35,0){\makebox(0,0){#3}}}

\newcommand{\elsta}[3]
{\multiput(0.06,0)(0.25,0){4}{\line(1,0){0.15}}
\put(1,0.35){\makebox(0,0){$#1$}}
\put(0,0){\circle*{0.1}}
\put(1,0){\circle*{0.1}}
\put(-0.35,0){\makebox(0,0){#2}}
\put(1.35,0){\makebox(0,0){#3}}}

\newcommand{\elstatNO}[3]
{\multiput(0.06,0.08)(0.01,0.01){14}{\tiny.}
\multiput(0.30,0.32)(0.01,0.01){14}{\tiny.}
\multiput(0.55,0.57)(0.01,0.01){14}{\tiny.}
\multiput(0.79,0.81)(0.01,0.01){14}{\tiny.}
\multiput(1.03,1.05)(0.01,0.01){14}{\tiny.}
\multiput(1.27,1.29)(0.01,0.01){14}{\tiny.}
\multiput(1.51,1.53)(0.01,0.01){14}{\tiny.}
\multiput(1.75,1.77)(0.01,0.01){14}{\tiny.}
\put(0.95,0.75){\makebox(0,0){\VectorUr}}
\put(0,0){\circle*{0.15}}
\put(2,2){\circle*{0.15}}
\put(0.75,1.25){\makebox(0,0){$#1$}}
\put(-0.5,0){\makebox(0,0){$#2$}}
\put(2.5,2){\makebox(0,0){$#3$}}
}

\newcommand{\elstatNW}[3]
{\multiput(-0.05,0.05)(-0.015,0.015){10}{\circle*{0.02}}
\multiput(-0.3,0.3)(-0.015,0.015){10}{\circle*{0.02}}
\multiput(-0.55,0.55)(-0.015,0.015){10}{\circle*{0.02}}
\multiput(-0.8,0.8)(-0.015,0.015){10}{\circle*{0.03}}
\multiput(-1.05,1.05)(-0.015,0.015){10}{\circle*{0.03}}
\multiput(-1.3,1.3)(-0.015,0.015){10}{\circle*{0.03}}
\multiput(-1.55,1.55)(-0.015,0.015){10}{\circle*{0.03}}
\multiput(-1.8,1.8)(-0.015,0.015){10}{\circle*{0.03}}
\put(-0.9,0.9){\makebox(0,0){\VectorUl}}
\put(0,0){\circle*{0.215}}
\put(-2,2){\circle*{0.2153}}
\put(-0.9,0.9){\makebox(0,0){\VectorUl}}
\put(0,0){\circle*{0.215}}
\put(-2,2){\circle*{0.215}}
\put(-0.5,1){\makebox(0,0){$#1$}}
\put(0,-0.5){\makebox(0,0){$#2$}}
\put(-2,2.5){\makebox(0,0){$#3$}}
}

\newcommand{\photonSE}[5]
{\put(0,0){\photonHS{#3}{#4}{}{}}
\put(1.5,0){\VPloopD{#1}{#2}}
\put(2,0){\photonHS{#5}{}{}{}}}

\newcommand{\photonSEt}[5]
{\put(0,0){\photonHS{#3}{#4}{}{}}
\put(1.5,0){\VPloopDt{#1}{#2}}
\put(2,0){\photonHS{#5}{}{}{}}}

\newcommand{\ElSE}[3]
{\qbezier(0,-1)(.2025,-1.1489)(0.3420,-0.9397)
\qbezier(0.3420,-0.9397)(0.4167,-0.7217)(0.6428,-0.766)
\qbezier(0.6428,-0.766)(0.8937,-0.7499)(0.866,-0.5)
\qbezier(0.866,-0.5)(0.7831,-0.2850)(0.9848,-0.1736)
\qbezier(0.9848,-0.1736)(1.1667,0)(0.9848,0.1736)
\qbezier(0.9848,0.1736)(0.7831,0.2850)(0.866,0.5)
\qbezier(0.866,0.5)(0.8937,0.7499)(0.6428,0.766)
\qbezier(0.6428,0.766)(0.4167,0.7217)(0.3420,0.9397)
\qbezier(0.3420,0.9397)(.2025,1.1489)(0,1)
\put(1.05,0.02){\VectorUp}
\put(0,1){\circle*{0.15}}
\put(0,-1){\circle*{0.15}}
\put(1.45,0){\makebox(0,0){$#1$}}
\put(-0.35,-1){\makebox(0,0){#2}}
\put(-0.35,1){\makebox(0,0){#3}}}

\newcommand{\ElSEL}[3]
{\qbezier(0,-1)(-.2025,-1.1489)(-0.3420,-0.9397)
\qbezier(-0.3420,-0.9397)(-0.4167,-0.7217)(-0.6428,-0.766)
\qbezier(-0.6428,-0.766)(-0.8937,-0.7499)(-0.866,-0.5)
\qbezier(-0.866,-0.5)(-0.7831,-0.2850)(-0.9848,-0.1736)
\qbezier(-0.9848,-0.1736)(-1.1667,0)(-0.9848,0.1736)
\qbezier(-0.9848,0.1736)(-0.7831,0.2850)(-0.866,0.5)
\qbezier(-0.866,0.5)(-0.8937,0.7499)(-0.6428,0.766)
\qbezier(-0.6428,0.766)(-0.4167,0.7217)(-0.3420,0.9397)
\qbezier(-0.3420,0.9397)(-.2025,1.1489)(0,1)
\put(-1,0.02){\VectorUp} 
\put(-1.45,0){\makebox(0,0){$#1$}}
\put(0.35,-1){\makebox(0,0){#2}} \put(0.35,1){\makebox(0,0){#3}}}

\newcommand{\SEpolt}[5]
{\qbezier(0,-1.5)(.2025,-1.6489)(0.3420,-1.4397)
\qbezier(0.3420,-1.4397)(0.4167,-1.2217)(0.6428,-1.266)
\qbezier(0.6428,-1.266)(0.8937,-1.2499)(0.866,-1)
\qbezier(0.866,-1)(0.7831,-0.7850)(0.9848,-0.6736)
\qbezier(1,-0.5)(1.1,-0.5)(0.9848,-0.6736)
\qbezier(1,0.5)(1.1,0.5)(0.9848,0.6736)
\qbezier(0.9848,0.6736)(0.7831,0.7850)(0.866,1)
\qbezier(0.866,1)(0.8937,1.2499)(0.6428,1.266)
\qbezier(0.6428,1.266)(0.4167,1.2217)(0.3420,1.4397)
\qbezier(0.3420,1.4397)(.2025,1.6489)(0,1.5)
\put(1,0){\VPloopLRt{#1}{#2}}
\put(0.87,-1){\VectorUp}
\put(0.67,1.23){\Vector}
\put(1.3,-1){\makebox(0,0){#3}}
\put(1,0.5){\circle*{0.15}}
\put(1,-0.5){\circle*{0.15}}
\put(0,1.5){\circle*{0.15}}
\put(0,-1.5){\circle*{0.15}}
\put(-0.35,-1.5){\makebox(0,0){#4}}
\put(-0.35,1.5){\makebox(0,0){#5}}}

\newcommand{\SEpoltNA}[5]
{\qbezier(0,-1.5)(.2025,-1.6489)(0.3420,-1.4397)
\qbezier(0.3420,-1.4397)(0.4167,-1.2217)(0.6428,-1.266)
\qbezier(0.6428,-1.266)(0.8937,-1.2499)(0.866,-1)
\qbezier(0.866,-1)(0.7831,-0.7850)(0.9848,-0.6736)
\qbezier(1,-0.5)(1.1,-0.5)(0.9848,-0.6736)
\qbezier(1,0.5)(1.1,0.5)(0.9848,0.6736)
\qbezier(0.9848,0.6736)(0.7831,0.7850)(0.866,1)
\qbezier(0.866,1)(0.8937,1.2499)(0.6428,1.266)
\qbezier(0.6428,1.266)(0.4167,1.2217)(0.3420,1.4397)
\qbezier(0.3420,1.4397)(.2025,1.6489)(0,1.5)
\put(1,0){\circle{1}}
\put(0.87,-1){\VectorUp}
\put(0.67,1.23){\Vector}
\put(1.3,-1){\makebox(0,0){#3}}
\put(1,0.5){\circle*{0.15}}
\put(1,-0.5){\circle*{0.15}}
\put(0,1.5){\circle*{0.15}}
\put(0,-1.5){\circle*{0.15}}
\put(-0.35,-1.5){\makebox(0,0){#4}}
\put(-0.35,1.5){\makebox(0,0){#5}}}

\newcommand{\SEpol}[5]
{\qbezier(0,-1.5)(.2025,-1.6489)(0.3420,-1.4397)
\qbezier(0.3420,-1.4397)(0.4167,-1.2217)(0.6428,-1.266)
\qbezier(0.6428,-1.266)(0.8937,-1.2499)(0.866,-1)
\qbezier(0.866,-1)(0.7831,-0.7850)(0.9848,-0.6736)
\qbezier(1,-0.5)(1.1,-0.5)(0.9848,-0.6736)
\qbezier(1,0.5)(1.1,0.5)(0.9848,0.6736)
\qbezier(0.9848,0.6736)(0.7831,0.7850)(0.866,1)
\qbezier(0.866,1)(0.8937,1.2499)(0.6428,1.266)
\qbezier(0.6428,1.266)(0.4167,1.2217)(0.3420,1.4397)
\qbezier(0.3420,1.4397)(.2025,1.6489)(0,1.5)
\put(1,0){\VPloopLR{#1}{#2}}
\put(0.87,-1){\VectorUp}
\put(0.67,1.23){\Vector}
\put(1.3,-1){\makebox(0,0){#3}}
\put(1,0.5){\circle*{0.15}}
\put(1,-0.5){\circle*{0.15}}
\put(0,1.5){\circle*{0.15}}
\put(0,-1.5){\circle*{0.15}}
\put(-0.35,-1.5){\makebox(0,0){#4}}
\put(-0.35,1.5){\makebox(0,0){#5}}}

\setlength{\unitlength}{0.6cm} \thicklines

\begin{abstract}
A formalism for energy-dependent many-body perturbation theory
(MBPT), previously indicated in our recent review articles
(Lindgren \it{et al.}, Phys.Rep. \textbf{389},161(2004),
Can.J.Phys. \textbf{83},183(2005)), is developed in more detail.
The formalism allows for a mixture of energy-dependent (retarded)
and energy-independent (instantaneous) interactions and hence for
a merger of QED and standard (relativistic) MBPT. This combination
is particularly important for light elements, such as light
heliumlike ions, where electron correlation is pronounced. It can
also be quite significant in the medium-heavy mass range, as
recently discussed by Fritzsche \it{et al.} (J.Phys.
B\textbf{38},S707(2005)), with the consequence that the effects
might be significant also in analyzing the data of experiments
with highly charged ions. A numerical procedure for treating the
combined effect is described, and some preliminary numerical
results are given for heliumlike ions. This represent the first
numerical evaluation of effects beyond two-photon exchange
involving a retarded interaction. It is found that for heliumlike
neon the effect of one retarded photon (with Coulomb interactions
of all orders) represents about 99\% of the non-radiative effects
beyond energy-independent MBPT.
\end{abstract}

\title{Many-body procedure for energy-dependent perturbation:
Merging many-body perturbation theory with QED}
\author{Ingvar Lindgren\footnote{e-mail
address: ingvar.lindgren@fy.chalmers.se}, Sten
Salomonson\footnote{e-mail address: f3asos@fy.chalmers.se}, and
Daniel Hedendahl\footnote{e-mail address: dahe@fy.chalmers.se}}
\affiliation{Physics Department, G\öteborg University, G\öteborg,
Sweden} \pacs{31.15.Md, 31.25.-v, 31.30.Jv}
\date{\today}
\maketitle

\section{Introduction}
What is commonly known as Many-Body Perturbation Theory (MBPT) is
a class of perturbative schemes for bound atomic, molecular or
nuclear states with a time- or energy-independent perturbation,
based upon the Rayleigh-Schr\ödinger perturbative scheme, such as
the Brueckner-Goldstone linked-diagram expansion or variants
thereof~\cite{LM86}. Also certain iterative or "all-order"
approaches, like the Coupled-Cluster Approach (CCA) or
"Exponential Ansatz", can be referred to this category, although
they are not strictly perturbative in nature. All these approaches
can in principle treat the electron correlation to arbitrary
order. In addition, they have---as distinct from procedures based
upon the Brillouin-Wigner (BW) perturbation expansion---the
important property of being \it{size extensive} in the sense that
the energy scales linearly with the size of the system.
Furthermore, by using an extended or multi-reference model space
such schemes can also successfully handle the \it{quasidegenerate}
problem with closely spaced energy levels that are strongly mixed
by the perturbation.

For time- or energy dependent perturbations, like those of
quantum-electrodynamics (QED), the situation is quite different
and much less developed. There is presently no numerical scheme
available that can treat energy-dependent perturbations together
with electron correlation to arbitrary order, and also the
treatment of quasidegeneracy forms a serious problem in connection
with such interactions. In the numerical methods presently
available for QED calculations the electron-electron interaction
is treated by the exchange of fully covariant photons, which is
quite a tedious---and usually unnecessarily tedious---process to
handle the electron correlation. At most two-photon exchange can
be treated in this way with computers available today, which is
insufficient for light and medium-heavy elements, where the
electron correlation is quite important.

A major problem in extending the energy-dependent perturbation
theory to include electron correlation is that most methods have a
structure that is quite different from that of energy-independent
perturbation theory, which makes it difficult to utilize the
well-developed methods of the latter. Of the available methods
only the \it{Covariant Evolution Operator} (CEO) method that we
recently developed, has a structure that is akin to standard
energy-independent MBPT~\cite{Li98,LAS01,LSA04}. This opens the
possibility of combining the two approaches as proposed in our
recent review articles~\cite{LSA04,LSH05} and further developed in
the present work. In that scheme the CEO method is combined with
all-order MBPT methods of coupled-cluster type, so that the
exchange of covariant photons can be mixed with an arbitrary
number of instantaneous Coulomb interactions. In this way electron
correlation can for the first time be treated to arbitrary order
together with energy-dependent interactions of QED type.

The CEO method was originally developed in order to be able to
treat the quasidegeneracy problem in QED calculations. The
standard procedure for bound-state QED is the $S$-matrix
formulation~\cite{MPS98}, which has been successfully applied
particularly to highly charged ions. For lighter elements---in
addition to the electron-correlation problem---also the
quasidegeneracy problem might be quite pronounced, and the
$S$-matrix formulation fails. One illustrative example is the fine
structure of heliumlike ions, where in the relativistic
formulation for instance the lowest triplet state $^3P_1$ is a
mixture of the basis states $1s2p_{1/2}$ and $1s2p_{3/2}$, which
are very closely spaced in energy and strongly mixed for light
elements. In order to be able to use the procedure of standard
MBPT, where quasidegenerate states are included in an extended
model space, off-diagonal elements of the effective Hamiltonian
have to be evaluated, which is not possible in the standard
$S$-matrix formulation, due to the energy conservation of the
scattering process~\cite{MS00}. In the CEO method, based upon the
evolution operator for \it{finite} time~\cite{LSA04}, the
extended-model-space technique can be used, and a few years ago we
applied this to evaluate QED contributions to the fine structure
for heliumlike ions, including the quasidegenerate $^3P_1$ state,
down to $Z=9$~\cite{LAS01}. The same problem has very recently
been treated by Shabaev \it{et al.} down to $Z=12$, using the
\it{two-times Green's function technique}~\cite{Shab02,AShab05}.
Furthermore, the presently available methods for numerical QED
calculations suffer from the shortcoming that they are not
applicable to the lightest elements---below Z=10, say---due to
convergence problems. The extension of the CEO method to include
electron correlation to arbitrary order, presented here, is
expected to remedy this problem.

The most accurate calculations on light heliumlike ions have been
performed by the analytical or "unified" method of Drake and
related techniques~\cite{Dr79,Dr88,Drake02,Pach02,PSap02}. This is
based upon expansion in powers of the fine-structure constant
$\alpha$ of the Bethe-Salpeter equation and the Brillouin-Wigner
(BW) perturbation series~\cite{Su57a,DK74}. It leads to extremely
high accuracy for the lightest elements, but there is still a
significant discrepancy between theory and experiment for the
lowest triplet state of neutral helium. Being based upon the BW
perturbation expansion, the procedure is \it{not} size-extensive
and less suitable for larger systems.

In addition to the light heliumlike ions and other light systems,
the combination of QED and many-body effects can be of importance
also for heavier systems, as recently discussed by Fritzsche
\it{et al.}~\cite{FIS05}. They have analyzed this problem
particularly with regard to possible heavy-ion experiments of the
type that can be performed at the big storage rings, like that at
GSI in Darmstadt. They then conclude that the accurate treatment
of the interplay between QED and many-body effects constitutes one
of the most challenging problems in connection with
highly-charged-ion experiments.

The present paper will be organized as follows. In the next
section we briefly review the standard perturbation theory for
time-independent and time-dependent perturbations, and next we
summarize the properties of our recently introduced covariant
evolution-operator method. The following main section deals with
the derivation of Bloch equations for combined retarded and
instantaneous interactions, which constitute our working equations
for treating this problem. Finally, we describe briefly our
numerical procedure and give some preliminary numerical results,
including the first numerical result of effects beyond two-photon
exchange with a retarded interaction. The numerical procedure
together with more complete numerical results will be published
separately~\cite{HSL06}.

\section{Standard Many-Body Perturbation Theory}
\subsection{Time-independent perturbation theory}
In the multi-reference form of MBPT we consider a number of
\it{target states} that are eigenstates of the Hamiltonian of the
system
\begin{equation}
  \label{SE}
  H\,\ket{\Psi^\alpha}=E^\alpha\,\ket{\Psi^\alpha}\,;\qquad
  (\alpha=1,2,\cdots d).
\end{equation}
The Hamiltonian is partitioned into a model Hamiltonian, $H_0$,
and a time-independent perturbation, $H'$,
\begin{equation}
  \label{Part}
  H=H_0+H'.
\end{equation}
For an $N$-electron system the model Hamiltonian is assumed to be
composed of single-electron Schr\ödinger or Dirac Hamiltonians
\begin{equation}
  \label{H0}
  H_0=\sum_i^N h_0(i).
\end{equation}
For each target state there is a model state, confined to a
subspace, the \it{model space}, with the projection operator $P$.
In the intermediate normalization we use here the model states are
the projection of the corresponding target states on the model
space
\begin{equation}
  \label{ZOWF}
  \ket{\Psi_0^\alpha}=P\ket{\Psi^\alpha}.
\end{equation}
A wave operator can be defined for the inverse transformation
\begin{equation}
  \label{Wop}
  \ket{\Psi^\alpha}=\Omega\,\ket{\Psi_0^\alpha}\,; \qquad
  (\alpha=1,2,\cdots d).
\end{equation}
An \it{effective Hamiltonian} can be defined, $H\eff=PH\Om P$,
that operates in the model space and for which the eigenvectors
are the model states and the eigenvalues the corresponding exact
energies
\begin{equation}
  \label{SecEq}
  H\eff\,\ket{\Psi_0^\alpha}=E^\alpha\,\ket{\Psi_0^\alpha}.
\end{equation}
The corresponding \it{effective interaction} is defined
\begin{equation}
  \label{EffInt}
  V\eff=H\eff-PH_0P=PH'\Omega P.
\end{equation}
The wave operator satisfies the generalized Bloch
equation~\cite{Bl58a,Bl58b,Li74}
\begin{equation}
  \label{Bloch}
  \boxed{\big[\Omega,H_0\big]P=\big(H'\Omega -\Omega\,
  V\eff\big)P}
\end{equation}
This leads to the Rayleigh-Schr\ödinger perturbative expansion for
a general multi-reference (quasidegenerate) model space, and it
can also be used to generate the corresponding linked-diagram
expansion of Brueckner-Goldstone type
\begin{equation}
  \label{BlochL}
  \big[\Omega,H_0\big]P=\big(H'\Omega -\Omega\,
  V\eff\big)_\linked P.
\end{equation}
Here, only so-called linked terms or diagrams survive on the
right-hand side. Using the exponential Ansatz
\begin{equation}
  \label{CCA}
  \Om=\{\me^S\},
\end{equation}
where the curly brackets represent normal-ordering, leads to the
Coupled-Cluster expansion for the same model space~\cite{Li78}
\begin{equation}
  \label{BlochC}
  \big[S,H_0\big]P=\big(H'\Omega -\Omega\,
  V\eff\big)_\conn P.
\end{equation}
Here, all terms on the right-hand side are "connected". (For the
distinction between "linked" and "connected", see, for instance,
ref.~\cite{LM86}.

\subsection{Time-dependent perturbation theory}
In this paper we shall only be concerned with \it{stationary
states}, but we need for our purpose to use the formalism for
time-dependent perturbation, which we shall briefly review.

The time-dependent state vector satisfies the time-dependent
Schr\ödinger equation (using relativistic units
$\hbar=c=m=\epsilon_0=1$)
\begin{equation}
  \label{TDSE}
  \im\Partder{t}\,\ket{\chi(t)}=H(t)\,\ket{\chi(t)},
\end{equation}
where
\begin{equation}
  \label{Ham}
  H(t)=H_0+H'(t).
\end{equation}
$H_0$ is the time-independent model Hamiltonian and $H'(t)$ is a
perturbation that might be time-dependent. In the \it{interaction
picture} (IP), where an operator is related to that in the
\it{Schr\ödinger picture }(SP)  by
\begin{equation}
  \label{IP}
  O_\mI(t)=\me^{\im H_0 t} O_\rm{S}\,\me^{-\im H_0 t}
\end{equation}
the Schr\ödinger equation becomes
\begin{equation}
  \label{SEIP}
  \im\Partder{t}\,\ket{\chi_\mI(t)}=H'_\mI(t)\,\ket{\chi_\mI(t)}.
\end{equation}
The \it{time-evolution operator} is defined by
\begin{equation}
  \label{EvolOp}
  \ket{\chi_\mI(t)}=U_\mI(t,t_0)\,\ket{\chi_\mI(t_0)}
\end{equation}
and satisfies the equation
\begin{equation}
  \im\Partder{t}\,U_\mI(t,t_0)=H'_\mI(t)\,U_\mI(t,t_0)
\end{equation}
with the solution~\cite{FW71}
\begin{equation}
  \label{Uexp}
  U_\mI(t,t_0)=1+\sum_{n=1}^\infty\frac{(-\im)^n}{n!}\int_{t_0}^t\dif^4x_n\cdots
  \int_{t_0}^t\dif^4x_1\;\TD\big[\calH'_\mI(x_n)\calH'_\mI(x_{n-1})
  \cdots \calH'_\mI(x_1)\big].
\end{equation}
Here,  $\TD$ is the Dyson time-ordering operator, and
$\calH'_\mI(x)$ is the perturbation \it{density}, defined by
\begin{equation}
  \label{calH}
  H'_\mI(t)=\intbx\,\calH'_\mI(t,\bx).
\end{equation}
An \it{adiabatic damping} is added to the perturbation
\begin{equation}
  \label{Damp}
  H'_\mI(t)\rarr H'_{\mI\gamma}=H'_\mI\,\me^{-\gamma\abs{t}}\,;\qquad
  U_\mI(t,t_0)\rarr U_{\gamma}(t,t_0),
\end{equation}
where $\gamma$ is a small, positive number. This implies that as
$t\rarr\pm\infty$ the eigenfunctions of $H$ tend to eigenfunctions
of $H_0$.

For stationary states we assume that the perturbation $H'$ is
time-independent in the Schr\ödinger picture---apart from the
adiabatic-damping factor. Then according to the Gell-Mann-Low
(GML) theorem~\cite{GML51} the wave function at time $t=0$ is for
a single-reference model space given by
\begin{equation}
  \label{GML0}
  {\ket{\Psi}}=\ket{\chi(0)}=
  \limgam
  \frac{\Ugam{0}\bigket{\Psi_0}}{\bra{\Psi_0}\Ugam{0}\ket{\Psi_0}},
\end{equation}
where $\ket{\Psi_0}$ is the time-independent model state. This can
be generalized to a general multi-reference model space~\cite[Eq.
110]{LSA04}
\begin{equation}
  \label{GML}
  {\ket{\Psi^\alpha}}=\limgam \frac{N^\alpha\Ugam{0}\ket{\Phi^\alpha}}
  {\bra{\Phi^\alpha}\Ugam{0}\ket{\Phi^\alpha}}\,;\qquad
  (\alpha=1,2,\cdots d),
\end{equation}
where $N^\alpha$ is a normalization factor and the vector
$\ket{\Phi^\alpha}$ is defined
\begin{equation}
  \label{Phi}
  \ket{\Phi^\alpha}=\limgam\lim_{t\rarr-\infty}\ket{\chi^\alpha(t)}.
\end{equation}
Then the wave function at time $t=0$ satisfies the
time-independent Schr\ödinger-like equation
\begin{equation}
  \label{Schr}
  \boxed{\big(H_0+H'\big){\Psi}^\alpha=E^\alpha{\Psi}^\alpha}
\end{equation}
where $H'$ is the time-independent perturbation in the
Schr\ödinger picture.

The evolution operator normally contains singularities or
quasi-singularities as $\gamma\rarr0$, when an intermediate state
is degenerate or closely degenerate with the initial state. In the
GML formulas these (quasi-)singularities are eliminated by the
denominator so that the ratio is always regular, which is one
formulation of the linked-diagram theorem~\cite{Go57}.

\section{Covariant evolution-operator approach}
(The reader is referred to refs~\cite{LSA04,LSH05} for more
details concerning the basic covariant evolution-operator
formalism.)

\subsection{Interaction with the electro-magnetic field}
\newcommand{\contract}[3]
{\begin{picture}(0,0.5)(#2,#3) \thinlines\put(0,0){\line(1,0){#1}}
\put(0,0){\line(0,-1){0.2}}\put(#1,0){\line(0,-1){0.2}}
\end{picture}}
We consider now the interaction between electrons and the
quantized electro-magnetic field represented by the perturbation
density~\cite{Sch61}
\begin{equation}
  \label{Int}
  \calH'(x)=-e\hpsi_\mI\dagg\alpha^\mu A_{\mu}\hpsi_\mI.
\end{equation}
Here, $\hpsi_\mI,\;\hpsi_\mI\dagg$ are the electron field
operators, $\alpha^\mu$ the Dirac alpha operators and $A_\mu$ is
the radiation field
\begin{equation}
  \label{Field}
  A_{\mu}\propto\eps^{r}_\mu \,\Big(a_r\dagg(\bk)\,
e^{i\kappa x}+a_r(\bk)\,e^{-i\kappa x}\Big),
\end{equation}
applying standard summation convention. $a_r\dagg(\bk),\,a_r(\bk)$
are the photon creation/absorption operators, $\bk$ is the wave
vector, $\kappa$ is the four-vector momentum
$\kappa=(\omega,-\bk)$ ($k=|\bk|$), and $\eps_{r}^\mu$ represent
the polarization vectors. The Hamiltonian of the radiation field
is represented by
\begin{equation}
  \label{RadHam}
  H_\rm{rad}=\omega\,a_r\dagg(\bk)a_r(\bk)
  =k\,a_r\dagg(\bk)a_r(\bk).
\end{equation}

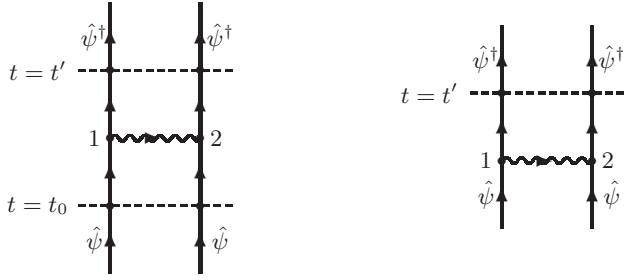
\begin{figure}[thb]
\begin{center}
\begin{picture}(7,6)(-1,-1.5)
\put(-0.75,3){\dash{14}} \put(-1.6,3){\makebox(0,0){$t=t'$}}
\put(0,3){\Elline{1.5}{0.75}{\hpsi\dagg\,}{}}
\put(2,3){\Elline{1.5}{0.75}{}{\,\hpsi\dagg}}
\put(0,0){\Elline{1.5}{0.75}{}{}}
\put(2,0){\Elline{1.5}{0.75}{}{}} \put(0,1.5){\photon{ }{1}{2}}
 \put(0,1.5){\Elline{1.5}{0.75}{}{}}
 \put(2,1.5){\Elline{1.5}{0.75}{}{}}
\put(0,3){\circle*{0.15}} \put(2,3){\circle*{0.15}}
 \put(-0.75,0){\dash{14}} \put(-1.6,0){\makebox(0,0){$t=t_0$}}
 \put(0,0){\circle*{0.15}} \put(2,0){\circle*{0.15}}
\put(0,-1.5){\Elline{1.5}{0.75}{\hpsi\,}{}}
\put(2,-1.5){\Elline{1.5}{0.75}{}{\,\hpsi}}
\end{picture}
\begin{picture}(7,6)(-2.5,-1)
\put(-0.75,3){\dash{14}} \put(-1.6,3){\makebox(0,0){$t=t'$}}
\put(0,3){\Elline{1.5}{0.75}{\hpsi\dagg\,}{}}
\put(2,3){\Elline{1.5}{0.75}{}{\,\hpsi\dagg}}
\put(0,0){\Elline{1.5}{0.75}{\hpsi\,}{}}
\put(2,0){\Elline{1.5}{0.75}{}{\,\hpsi}}
\put(0,1.5){\photon{}{1}{2}}
 \put(0,1.5){\Elline{1.5}{0.75}{}{}}
 \put(2,1.5){\Elline{1.5}{0.75}{}{}}
\put(0,3){\circle*{0.15}} \put(2,3){\circle*{0.15}}
\end{picture}
    \renewcommand{\normalsize}{\footnotesize}
    \caption{Graphical representation of the covariant-evolution
      operator for single-photon exchange in the form \eq{CovEv} (left) and in the
      form \eq{CovEv1} with $t_0\rarr-\infty$.}
    \renewcommand{\normalsize}{\standard}
    \label{Fig:SingPhot}
\end{center}
\end{figure}

With the interaction density \eqref{Int} the evolution operator
\eqref{Uexp} for single-photon exchange becomes
\begin{equation}
  \label{NoncovEv}
  U\tva(t',t_0)=-\half\dint_{-t_0}^{t'}\dif^4x_1\dif^4x_2\,
  \hpsi_{\mI+}\dagg(x_1)\hpsi_{\mI+}\dagg(x_2)\,
  \im I(x_1,x_2)\,\hpsi_{\mI+}(x_2)\hpsi_{\mI+}(x_1),
\end{equation}
integrated over all space coordinates and time coordinates as
indicated. $\hpsi_{\mI+},\,\hpsi_{\mI+}\dagg$ are the
positive-energy part of the electron-field operators, and the
interaction kernel
\begin{equation}
\label{Interact}\im I(x_1,x_2)
=\contract{2.7}{-2}{-0.8}(-e\alpha^\mu A_\mu)_1
  (-e\alpha^\nu A_\nu)_2
  =e^2\alpha_1^{\mu}\alpha_2^{\nu}\,\im D   _{F\mu\nu}(x_1-x_2)
\end{equation}
is given by the product of two perturbations \eqref{Int} with
contraction of the radiation-field operators (indicated by the
hook). $\DF(x_1-x_2)$ is the Feynman \it{photon propagator}.

The evolution operator above is \it{non-covariant} but can be made
\it{covariant} by inserting zeroth-order Green's functions on the
in- and outgoing states~\cite{LAS01,LSA04}
\begin{eqnarray}
  \label{CovEv}
  U\tva_\Cov(t',t_0)&=&-\half \dint\dif^3\bx'_1\dif^3\bx'_2\, \hpsi_\mI\dagg(x'_1)\hpsi_\mI\dagg(x'_2)
  \dint\dif^4x_1\dif^4x_2\,G_0(x_1',x_2';x_1,x_2)\,\nn
  &\times&\dint\dif^3\bx_{10}\dif^3\bx_{20}\,\im I(x_1,x_2)\,G_0(x_1,x_2;x_{10},x_{20})
  \,\hpsi_\mI(x_{20})\hpsi_\mI(x_{10}).
\end{eqnarray}
Here, the time integrations over $t_1$ and $t_2$ are performed
over \it{all times}, and positive- as well as negative-energy
states are allowed as incoming and outgoing states. The initial
and final times are the same for the two electrons, i.e.,
  \[t_{10}=t_{20}=t_0\quad \rm{and}\quad t'_1=t'_2=t'.\]
We shall in the following assume that the initial time is
$t_0=-\infty$, and then due to the adiabatic damping \eqref{Damp}
we can leave out the rightmost Green's function~\cite{LSA04}
\begin{equation}
  \label{CovEv1}
  U\tva_\Cov(t',-\infty)=
  \half \hpsi_\mI\dagg(x'_1)\hpsi_\mI\dagg(x'_2)\,G_0(x_1',x_2';x_1,x_2)\,\im I(x_1,x_2)
  \,\hpsi_\mI(x_{2})\hpsi_\mI(x_{1}).
\end{equation}
Here, we have left out the integrations, and in the following we
shall also leave out the subscript $_\Cov$ as well as the initial
time.

\subsection{Wave operator and effective interaction}

\begin{figure}[htb]
\setlength{\unitlength}{0.8cm}
\begin{center}
\begin{picture}(5,4)(-2,0)
\put(0,1){\Elline{2}{1.5}{r}{}} \put(2,2){\Elline{1}{0.5}{}{s}}
 \put(0,1){\photonENE{k}{}{}}
 \put(0,0){\Elline{1}{0.5}{t}{}}
 \put(2,0){\Elline{2}{0.5}{}{u}}
\end{picture}
\begin{picture}(5,4)(-2,0)
\put(2,1){\Elline{2}{1.5}{}{s}} \put(0,2){\Elline{1}{0.5}{r}{}}
 \put(2,1){\photonWNW{k}{}{}}
 \put(2,0){\Elline{1}{0.5}{}{u}}
 \put(0,0){\Elline{2}{0.5}{t}{}}
\end{picture}
    \renewcommand{\normalsize}{\footnotesize}
    \caption{Graphical representation of single-photon potential \eqref{VExpl}.}
    \renewcommand{\normalsize}{\standard}
    \label{Fig:SPPot}
\end{center}
\end{figure}
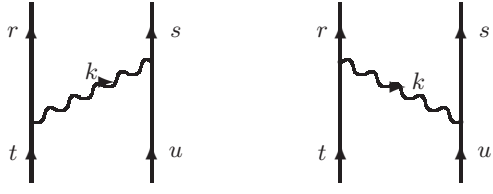

The evolution operator is generally singular and can be
expressed~\cite{LSA04}
\begin{equation}
  \label{Ured0}
  U(t)P=P+\Util(t)P\bsdot PU(0)P,
\end{equation}
where $\Util(t)$ is always regular and known as the \it{reduced
evolution operator}---all singularities are collected in the last
factor $PU(0)P$. The heavy dot indicates here that the two factors
evolve in time independently from different model-space states.
For the time $t=0$ this becomes
\begin{equation}
  \label{Ured}
  U(0)P=\big[1+Q\Util(0)\big]P\bsdot PU(0)P,
\end{equation}
where $Q$ is the projection operator for the "complementary space"
(outside the model space).

Inserting the expression \eqref{Ured} into the GML formula
\eqref{GML}, yields
\begin{equation}
  \label{}
  {\ket{\Psi^\alpha}}=\big[1+Q\Util(0)\big]P\limgam \frac{N^\alpha\Ugam{0}\ket{\Phi^\alpha}}
  {\bra{\Phi^\alpha}\Ugam{0}\ket{\Phi^\alpha}}
  =\big[1+Q\Util(0)\big]P{\ket{\Psi^\alpha}}.
\end{equation}
But $P{\ket{\Psi^\alpha}}=\Psi_0^\alpha$ is the model state
\eqref{ZOWF}, and hence the expression in the square brackets
represents the wave operator \eqref{Wop}
\begin{eqnarray}
  \label{WaveOp}
  \boxed{\Om=1+Q\Util(0)}
\end{eqnarray}

The $Q$ operator is for a two-electron system given by
\begin{equation}
  \label{Q}
  Q=1-P=\ket{rs}\bra{rs},
\end{equation}
where the ket vectors $\ket{rs}$ represent straight
(non-antisymmetrized) products of single particle states, summed
over all states outside the model space. The single-particle
states are generated by the single-particle Hamiltonians
\eqref{H0}
\begin{eqnarray}
  \label{OrbEq}
  h_0\ket{i}=\eps_i\ket{i},
\end{eqnarray}
which also include the nuclear field (Furry picture). In the
two-electron system we shall study here, the occupied electron
states are treated as open-shell or valence states, which implies
that there are no core or hole states---apart from the
negative-energy states.

Performing the integrations for the single-photon exchange
\eqref{CovEv1}, yields~\cite{LAS01,LSA04}
\begin{equation}
  \bra{rs}U\tva(t)\ket{ab}=\bigbra{rs}\me^{-\im t(E_{ab}-\eps_r-\eps_s)}
  \GQ(E_{ab})\V_1(E_{ab})\bigket{ab}\,;\qquad E_{ab}=\eps_a+\eps_b
\end{equation}
or, generally, operating on a model-space state of energy $\E$,
\begin{equation}
  \label{U2}
  U\tva(t)P=\me^{-\im t(\E-H_0)}
  \GQ(\E)\V_1(\E) P.
\end{equation}
Here,
\begin{equation}
  \label{GammaQ}
  \GQ(\E)=\frac{Q}{\E-H_0}
\end{equation}
is the \it{resolvent}, and
\begin{equation}
  \label{VExpl}
  \bra{rs}\V_1(\E)\ket{tu}=\Bigbra{rs}\int\dif k\,f(\bx_1,\bx_2,k)\Big[\frac{1}
  {\E-\eps_r-\eps_u-(k-\im\gamma)_r}
  +\frac{1}{\E-\eps_s-\eps_t-(k-\im\gamma)_s}\Big]\Bigket{tu}
\end{equation}
is the matrix element of the potential, considering both time
orderings (see Fig. \ref{Fig:SPPot}). The subscript $_r$
represents the sign of $\eps_r$. Using the evolution operator
\eqref{U2} and the relation \eqref{WaveOp} (with
$\Util\tva=U\tva$), this gives the wave operator for single-photon
exchange
\begin{equation}
  \label{WOett}
  \Om\ett P=\GQ(\E)\V_1(\E)P.
\end{equation}

The effective interaction \eqref{EffInt} can generally be
expressed~\cite{LSA04}
\begin{equation}
  \label{EffInt2}
  \boxed{V\eff=P\Big[\im\Partder{t}\Util(t)\Big]_{t=0}P}
\end{equation}
The reduced evolution operator has the same time dependence in all
orders as in first order \eqref{U2}, which implies that the time
derivation eliminates the resolvent. In first order this yields
\begin{equation}
  \label{Heffett}
  V\eff\ett(\E)=P\V_1(\E)\,P.
\end{equation}

The function $f(\bx_1,\bx_2,k)$ in the potential \eqref{VExpl}
depends on the gauge used and is in the Feynman gauge given by
\begin{equation*}
  f_\rm{F}(\bx_1,\bx_2,k)=-\frac{e^2}{4\pi^2\,}\,(1-\bs{\alpha}_{1}\bdot\bs{\alpha}_{2})
  \,\frac{\sin(kr_{12})}{\rr},
\end{equation*}
where $r_{12}$ is the interelectronic distance.

In the Coulomb gauge, which is natural to use in many-body
calculations, the potential can be separated into an
\it{instantaneous} and a \it{retarded} part,
\begin{equation}
  \label{V}
  V(\E)=V_\mI+V_\rm{Ret}(\E),
\end{equation}
where  only the latter is energy dependent. The instantaneous part
is the Coulomb interaction
\begin{equation}
  \label{VI}
  V_\mI=V_{12}=\frac{e^2}{4\pi\rr},
\end{equation}
and the retarded part is given by the expression \eq{VExpl} with
\begin{equation}
  \label{VExplC}
  f_\rm{C}(\bx_1,\bx_2,k)=
  \frac{e^2}{4\pi^2}\Big[-\bs{\alpha_1\cdot\alpha_2}
  \frac{\sin(k\rr)}{\rr}+(\bs{\alpha_1\cdot\nabla_1)}\,
  (\bs{\alpha_2\cdot\nabla_2}) \frac{\sin(k\rr)}{k^2\,\rr}\Big],
\end{equation}
where the nabla operators do not operate beyond the square
bracket. Here, the first term represent the \it{Gaunt} part and
the second term the \it{scalar-retardation} part, which together
form the \it{Breit interaction}. In the following we shall assume
that the Coulomb gauge is used.

The relations given here, particularly the framed equations
\eqref{WaveOp} and \eqref{EffInt2}, demonstrate the close analogy
between the covariant evolution-operator approach and standard
MBPT, which opens up the possibility for a merger of the two
procedures.

\section{Bloch equation for instantaneous and retarded interactions}
\subsection{Retarded interactions}

From the definition \eqref{Ured0}, the following relation can be
derived, using standard algebra~\cite{LSH05},
\begin{equation}
  \label{OmRec2}
  \Util(t) P=\Ubar(t) P+\Util(t)\,\big(P\Ubar P-\bsdot P\Ubar
  P\big),
\end{equation}
where
\begin{equation}
  \label{Ubar}
  \Ubar(t) P=\Util(t) P-M(t)
\end{equation}
is the evolution operator without intermediate model-space states
and $M$ is the \it{model-space contribution} (MSC). The $\Ubar$
operator in Eq. \eqref{OmRec2} has the time argument $t=0$.

As mentioned, in the product $\Util(t)P\Ubar(0)P$ the
time-dependent part $\Util(t)$ evolves from the energy of the
state to the far right ($\E$), while in the dot product
$\Util(t)\bsdot P\Ubar(0)P$ the operator $\Util(t)$ evolves from
the energy of the intermediate state ($\E'$). Therefore, the
second term, which represents the MSC part, becomes
\begin{equation}
  \label{MSC1}
  M=\big(\Util(\E)-\Util(\E')\big)\,P\Ubar P.
  \end{equation}
This can be expressed
\[P\Ubar P=\frac{\Hbar\eff}{\E-\E'},\]
where $\Hbar\eff$ is the analogue of the effective interaction
\eqref{EffInt2} without MSC
\begin{equation}
  \label{HBar}
  \Hbar\eff=P\Big[\im\Partder{t}\Ubar(t)\Big]_{t=0}P.
\end{equation}
This yields
\begin{equation}
  \label{MSC}
  M=\frac{\Util(\E)-\Util(\E')}{\E-\E'}\;\Hbar\eff
  =\partdelta{\Util}{\E}\;\Hbar\eff,
\end{equation}
using the difference ratio defined in Appendix \ref{sec:Diff}.
This yields
\begin{equation}
  \label{Om1}
  \Util(t) P=\Ubar(t)P+\pd{\Ubar}\,\Hbar\eff.
\end{equation}
With the wave-operator relation \eqref{WaveOp} we then have
\begin{equation}
  \label{Om1a}
  \Om P=\Ombar P+\pd{\Om}\,\Hbar\eff,
\end{equation}
where the last term represents the MSC (including "folded"
diagram) and $\Ombar$ is the wave operator without MSC.

The relation \eqref{OmRec2} leads to the expansion
\begin{equation}
  \label{OmRec3}
  \Util(t) P=\Ubar(t) P+\Ubar(t)\,\big( P\Ubar P-\bsdot P\Ubar P\big)
  +\Ubar(t)\,\big( P\Ubar P-\bsdot P\Ubar P\big)
  \big( P\Ubar P-\bsdot P\Ubar P\big)+\cdots
\end{equation}
The exchange of a single photon corresponds to the second-order
evolution operator and leads to the result given above
\eqref{WOett}. The two-photon exchange corresponds to the next
even order of the evolution operator
\begin{equation}
  \label{Om2}
  \Util\fyr(t) P=\Ubar\fyr(t) P+\Ubar\tva(t)\,
  \big( P\Ubar\tva P-\bsdot P\Ubar\tva P\big).
\end{equation}
Since there is no MSC in lowest order, we have
$U\tva=\Ubar\tva=\Util\tva$, and the corresponding wave operator
becomes
\begin{equation}
  \label{Om2a}
  \Omega\tva P=Q\Util\fyr(0)P
  =\Ombar\tva P+\partdelta{\Omega\ett}{\E}\;V\eff\ett,
\end{equation}
where $\Ombar\tva P=\GQ\V_1\GQ\V_1 P$. In the case of degeneracy
this goes over into
\begin{equation}
  \label{Om2b}
  \Omega\tva P=\Ombar\tva P+\partder{\Omega\ett}{\E}\;V\eff\ett.
\end{equation}

In third order we have~\cite[App. D]{LSH05}
\begin{equation}
  \label{Om3a}
  \Omega\tre P=\Ombar\tre P+\pd{\Om\tva}\;V\eff\ett+\pd{\Omega\ett}\;\Hbar\eff\tva,
\end{equation}
which leads to
\begin{equation}
  \label{Om3}
  \Omega\tre P=\Ombar\tre P+\pd{\Omega\ett}\;V\eff\tva
  +\pd{\Ombar\tva}\;V\eff\ett+\pdn{2}{\Omega\ett}\;\big(V\eff\ett\big)^2,
\end{equation}
where the second-order difference ratio is defined in Appendix
\ref{sec:Diff}. The last term is associated with double
model-space contributions ("double fold"). This leads to
conjecture for the all-order expansion (c.f.~\cite[Eq.116]{LSH05})
\begin{equation}
  \label{Key}
  \boxed{\Om P=\Ombar P+\sum_{n=1}^\infty
 \partdeltan{n}{\Ombar}{\E}\;\big(V\eff\big)^{n}}
 \end{equation}
which we shall now verify.

In order to prove the relation above, we start by using this
relation to form the difference ratio
\begin{eqnarray}
  \label{OmDiff1}
  \pd{\Om}&=&\pd{\Ombar}+\pdn{2}{\Ombar}\,V\eff+\pdn{3}{\Ombar}\,(V\eff)^2
  +\cdots+\pd{\Ombar}\pd{V\eff}+\pdn{2}{\Ombar}\,V\eff\pd{V\eff}+\cdots\nn
  &=&\sum_{n=1}^\infty
  \pdn{n}{\Ombar}\;(V\eff)^{(n-1)}\Big[1+\pd{V\eff}\Big].
\end{eqnarray}
(It should be noted that the different $V\eff$ operators are in
general associated with different energies, as explained in
Appendix \ref{sec:Diff}, Eq. \ref{DiffVeff}.) Next, we take the
time derivative of the relation \eqref{Om1}, using the relation
\eqref{EffInt2}, which yields
\begin{equation}
  \label{H1}
  V\eff=\Big[1+\pd{V\eff}\Big]\Hbar\eff.
\end{equation}
This gives with the relation \eqref{OmDiff1}
\begin{equation}
  \label{OmDiff2}
  \pd{\Om}\,\Hbar\eff=\sum_{n=1}^\infty
  \pdn{n}{\Ombar}\;(V\eff)^{n},
\end{equation}
and with the relation \eqref{Om1} we retrieve the relation
\eqref{Key}, which is then proven.

In order to obtain more Bloch-like relations, we first introduce
the \it{reaction operator}, which is the effective interaction
\eqref{EffInt2}, apart from the projection operators,
\begin{equation}
  \label{ReactOp}
  \VR=\Big[\im\Partder{t}\Util(t)\Big]_{t=0}.
\end{equation}
This gives $V\eff=P\VR P$, and the reaction operator is equal to
the wave operator, apart from the resolvent,
\begin{equation}
  \label{ReactOp2}
  Q\Util(0)=Q\Om=\GQ\VR.
\end{equation}
We also introduce the operator
\begin{equation}
  \label{ReactOpBar}
  \VRbar=\Big[\im\Partder{t}\Ubar(t)\Big]_{t=0},
\end{equation}
where $\Ubar$ is the evolution operator without MSC. This leads in
analogy with the relation \eqref{ReactOp2} to
\begin{equation}
  \label{ReactOpBar2}
  Q\Ubar(0)=Q\Ombar=\GQ\VRbar.
\end{equation}
Using the rule for differentiating a product, developed in
Appendix \ref{sec:Diff} (Eq. \ref{Diffn}), and the relation
\begin{equation}
  \label{GammaDer}
  \pdn{m}{\GQ}=-\GQ\pdn{{(m-1)}}{\GQ},
\end{equation}
we can express the difference ratios of $\Ombar$ as
\begin{equation}
  \label{OmBarDer}
  \pdn{n}{\Ombar}=\pdn{n}{(\GQ\VRbar)}=
  \sum_{m=0}^n\pdn{m}{\GQ}\,\pdn{{(n-m)}}{\VRbar}
  =\GQ\pdn{n}{\VRbar}-\sum_{m=1}^n\GQ\pdn{{(m-1)}}{\GQ}\,\pdn{{(n-m)}}{\VRbar}.
  \end{equation}
The last term can be expressed
\begin{equation}
  -\GQ\sum_{m=0}^{n-1}\pdn{m}{\GQ}\,\pdn{{(n-m)}}{\VRbar}=-\GQ\pdn{{(n-1)}}{\Ombar}.
\end{equation}
Inserted in the expansion \eqref{Key}, this yields
\begin{equation}
  \label{116}
  \boxed{Q\Om P=Q\Ombar
  P-\GQ\Om
  V\eff+\GQ\sum_{n=1}^\infty\pdn{n}{\VRbar}\,(V\eff)^n}
  \end{equation}
This equation will later be used to derive the Bloch equations for
energy-dependent interactions.

\subsection{Instantaneous interactions}

\begin{figure}[thb]
\begin{center}
\begin{picture}(3.5,4)(-1.5,0)
\put(-1.5,1.5){\makebox(0,0){\Large$\rho_{ab}$\;=}}
\put(0,1.5){\Elline{1.5}{0.75}{r}{}}
\put(1.5,1.5){\Elline{1.5}{0.75}{}{s}}
\put(0,0){\Elline{1.5}{0.75}{a}{}}
{\linethickness{1mm}\put(0,1.5){\line(1,0){1.5}}}
\put(1.5,0){\Elline{1.5}{0.75}{}{b}}
\end{picture}
\begin{picture}(3.5,4)(-1.5,0)
\put(-1,1.5){\makebox(0,0){\Large$=$}}
\put(0,0){\Elline{3}{1.5}{a}{}} \put(1.5,0){\Elline{3}{1.5}{}{b}}
\end{picture}
\begin{picture}(3.5,4)(-1.5,0)
\put(-1,1.5){\makebox(0,0){\Large$+$}}
\put(0,2){\Elline{1}{0.5}{r}{}}
\put(1.5,2){\Elline{1}{0.5}{}{s}}\put(0,1){\Elline{1}{0.5}{t}{}}
\put(1.5,1){\Elline{1}{0.5}{}{u}} \put(0,0){\Elline{1}{0.5}{a}{}}
\put(0,2){\elstatH{}{}{}}
{\linethickness{1mm}\put(0,1){\line(1,0){1.5}}}
\put(1.5,0){\Elline{1}{0.5}{}{b}}
\end{picture}
\begin{picture}(3.5,4)(-2,0)
\put(-1.5,1.5){\makebox(0,0){\Large$-$}}
\put(0,2.5){\Elline{0.75}{0.5}{r}{}}
\put(1.5,2.5){\Elline{1}{0.5}{}{s}}
\put(0,1.5){\Elline{1}{0.5}{c\;}{}}
\put(1.5,1.5){\Elline{1}{0.5}{}{\;d}}
\put(0,0){\Elline{0.75}{0.4}{a}{}}
{\linethickness{1mm}\put(0,2.5){\line(1,0){1.5}}}
\put(1.5,0){\Elline{0.75}{0.4}{}{b}}
\put(-0.25,0.75){\Ebox{2}{0.8}}\put(0.75,1.15){\makebox(0,0){$V\eff$}}
\end{picture}\\\begin{picture}(3.5,5)(0.3,-0.5)
\put(-1,1.5){\makebox(0,0){\Large$=$}}
\put(0,0){\Elline{3}{1.5}{a}{}} \put(1.5,0){\Elline{3}{1.5}{}{b}}
\end{picture}
\begin{picture}(3.5,4)(0.3,-0.5)
\put(-1,1.5){\makebox(0,0){\Large$+$}}
\put(0,1.5){\Elline{1.5}{0.75}{r}{}}
\put(1.5,1.5){\Elline{1.5}{0.75}{}{s}}
\put(0,0){\Elline{1.5}{0.75}{a}{}} \put(0,1.5){\elstatH{}{}{}}
\put(1.5,0){\Elline{1.5}{0.75}{}{b}}
\end{picture}
\begin{picture}(3.5,4)(0.3,-0.5)
\put(-1,1.5){\makebox(0,0){\Large$+$}}
\put(0,2){\Elline{1}{0.5}{r}{}} \put(1.5,2){\Elline{1}{0.5}{}{s}}
\put(0,1){\Elline{1}{0.5}{}{}} \put(1.5,1){\Elline{1}{0.5}{}{}}
\put(0,0){\Elline{1}{0.5}{a}{}} \put(0,2){\elstatH{}{}{}}
\put(0,1){\elstatH{}{}{}} \put(1.5,0){\Elline{1}{0.5}{}{b}}
\end{picture}
\begin{picture}(3.5,4)(0.3,0)
\put(-1,2){\makebox(0,0){\Large$+$}}
\put(0,3){\Elline{1}{0.5}{r}{}} \put(1.5,3){\Elline{1}{0.5}{}{s}}
\put(0,2){\Elline{1}{0.5}{}{}} \put(1.5,2){\Elline{1}{0.5}{}{}}
\put(0,1){\Elline{1}{0.5}{}{}} \put(1.5,1){\Elline{1}{0.5}{}{}}
\put(0,0){\Elline{1}{0.5}{a}{}} \put(0,3){\elstatH{}{}{}}
\put(0,2){\elstatH{}{}{}} \put(0,1){\elstatH{}{}{}}
\put(1.5,0){\Elline{1}{0.5}{}{b}}
\put(4.5,2){\makebox(0,0){\Large$+\cdots+\rm{folded}$}}
\end{picture}
    \renewcommand{\normalsize}{\footnotesize}
    \caption{Graphic representation of pair functions $\rho_{ab}$ generated by the
      pair equation (\ref{eq:PairEquation}).}
      \renewcommand{\normalsize}{\standard}
    \label{Fig:PairEqn}
  \end{center}
\end{figure}
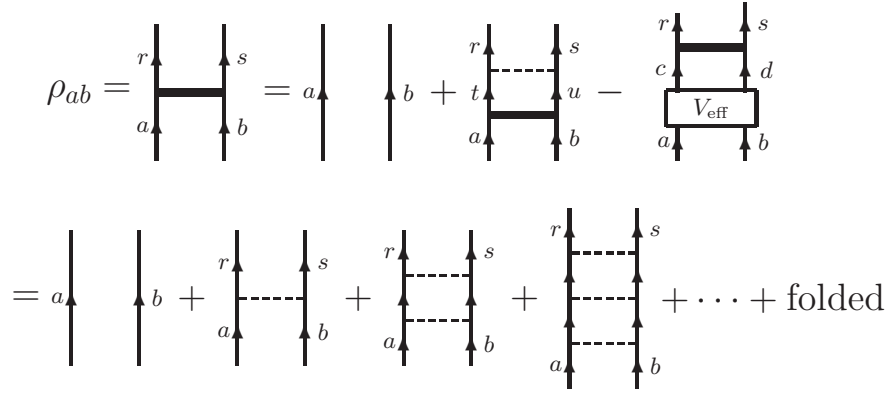

The instantaneous Coulomb interaction can be treated essentially
as in standard (relativistic) many-body theory, and we start by
recalling the treatment of the correlation effect for a
two-electron system. The wave operator in the coupled-cluster
formalism \eqref{CCA} can then be expressed~\cite{LM86}
\begin{eqnarray}
  \label{eq:WaveOpenShell}
  \Omega=1+S_{2},
\end{eqnarray}
where $S_2$ is the two-body cluster operator. Operating on a
model-space state, yields a \it{pair function}
\begin{eqnarray}
  \label{PairEq}
  \ket{\rho_{ab}}=\Om\ket{ab}=\ket{ab}+s^{rs}_{ab}\ket{rs}.
\end{eqnarray}
Inserting the pair function into the Bloch equation
\eqref{BlochC}, yields the corresponding pair equation
\begin{eqnarray}
  \label{eq:PairEquation}
  \big(\eps_{a}+\eps_{b}-h_0(1)-h_0(2)\big)\ket{\rho_{ab}}=
  \ip{rs}V_{12}\ket{\rho_{ab}}-
  \ket{\rho_{cd}}\bra{cd}V\eff\ket{ab}.
\end{eqnarray}
The last term is the \it{folded} term and is the result of the
reduction of singularities which appear when the intermediate
states lie in the model space, which we have referred to above as
\it{model-space contribution} (MSC). When the equation is solved
iteratively, the Coulomb interactions are generated to all orders,
as illustrated in Fig. \ref{Fig:PairEqn}. This is the type of pair
functions we have been using in our many-body calculations for
several decades~\cite{Ma79,Li85,ELi88,SO90,SY91}.

We denote the wave operator with only Coulomb interactions by
$\Om_\mI$ and the part with no folded diagrams  by $\Ombar_\mI$.
Then we have
\begin{equation}
  \label{OmbarI1}
  \Ombar_\mI P=\Big[1+\GQ V_{12}+\GQ V_{12}\GQ V_{12}
  +\cdots\Big]P,
\end{equation}
and using the relations \eqref{116} and \eqref{Key}, this leads to
the standard Bloch equation \eqref{Bloch}
\begin{equation}
  \label{BlochI}
  \boxed{\big[\Om_\mI,H_0\big]P=V_{12}\Om_\mI P-\Om_\mI\,V\eff}
\end{equation}

\subsection{Combined instantaneous and retarded interactions}
\renewcommand{\v}{H'}

We shall now find Bloch equations for the \it{combined} retarded
and instantaneous interactions. In the Coulomb gauge the function
$f(\bx_1,\bx_2,k)$, involved in the exchange of a retarded photon,
is given by the expression \eqref{VExplC}, which can be separated
into products of single-electron operators, as shown in Appendix
\ref{App:Coul},
\begin{equation}
  \label{fk}
  f_\rm{C}(\bx_1,\bx_2,k)=\frac{e^2k}{4\pi^2}\sum_{l=0}^\infty\Big[
  -(2l+1)V_\G^l(kr_1)\bsdot V_\G^l(kr_2)
  +\frac{1}{2l+1}V_{\sr}^l(kr_1)\bsdot V_{\sr}^l(kr_2)\Big].
\end{equation}
The two terms represent the Gaunt and scalar-retardation parts,
respectively, and we shall treat each of them as the result of
\it{two} perturbations, namely $V_\G^l(kr_1)$ and $V_\G^l(kr_2)$
in the case of the Gaunt interaction and $V_{\sr}^l(kr_1)$ and
$V_{\sr}^l(kr_2)$ for the scalar retardation---of course, with the
appropriate factors and with summation over the angular momentum
of the photon, $l$, and integration over space and the linear
photon momentum, $k$.

The photon can also be absorbed by the same electron, leading to
self-energy and vertex-correction contributions, as we shall
briefly indicate below.

The perturbations above are time-independent in the Schr\ödinger
picture, and we can then apply the Gell-Mann--Low theorem
\eqref{GML}, which leads to the Schr\ödinger-like equation
\eqref{Schr}
\begin{equation}
  \label{SCHR}
  \big(\bH_0+H'\big)\bs{\Psi}^\alpha=E^\alpha\bs{\Psi}^\alpha,
\end{equation}
where $H'$ is given by the instantaneous Coulomb interaction
$V_\mI=\VI$ and the two retarded components $V_\G^l(kr)$ and
$V_{\sr}^l(kr)$. The wave function lies here in an \it{extended
Fock space} with variable number of uncontracted, virtual photons,
which we indicate by using the bold-face symbol
$\bs{\Psi}^\alpha$. The bold-face symbol $\bH_0$ represents the
model Hamiltonian including the radiation field \eqref{RadHam}. We
also introduce a \it{Fock-space wave operator} in analogy with the
standard wave operator \eqref{Wop}
\begin{equation}
  \label{Wop1}
  \bs{\Psi}^\alpha=\OM\bs{\Psi}_0^\alpha.
\end{equation}

The resolvent \eqref{GammaQ} is  now generalized to
\begin{equation}
  \label{GammaQ1}
  \bGQ(\E)=\frac{\Q}{\E-\bH_0}
\end{equation}
when operating on a model-space state of energy $\E$. $\Q=1-P$ is
here the projection operator for the complementary Fock space, for
a two-electron system given by
\begin{equation}
  \label{QQ}
  \Q=\ket{rs}\bra{rs}+\ket{ij,k}\bra{ij,k}+\cdots
\end{equation}
The first term represents the part of the operator in the
restricted space with no photons (c.f. Eq. \ref{Q}) with
$\ket{rs}$ being a state outside the model space. The second term
represents the part with one photon, with $\ket{ij}$ being an
arbitrary state, etc.

\begin{figure}[hbt]
\begin{center}
\setlength{\unitlength}{0.6cm}
\begin{picture}(4,4.5)(-0.75,0)
\put(-2.3,2.25){\makebox(0,0){\large$\rho_{\G,ab}^l(k)$\;=}}
\put(3.1,2.25){\makebox(0,0){\LARGE=}}
\put(0,1.5){\Elline{2.5}{2}{r\;}{}}
\put(2,1.5){\Elline{2.5}{2}{}{\;s}} \put(0,2){\PhotonNe{k\;}{}{}}
\put(0,0.5){\Elline{1}{0.5}{a\;}{}}
\put(2,0.5){\Elline{1}{0.5}{}{\;b}}
\end{picture}
\begin{picture}(4,4.5)(-0.75,0)
\put(3.1,2.25){\makebox(0,0){\LARGE+}}
\put(0,1.5){\Elline{2.5}{2}{r\;}{}}
\put(2,1.5){\Elline{2.5}{2}{}{\;s}}
\put(0,2.5){\photonNe{k\;}{}{}}
\put(0,0.5){\Elline{1}{0.5}{a\;}{}}
{\linethickness{1mm}\put(0,1.75){\line(1,0){2}}}
\put(2,0.5){\Elline{1}{0.5}{}{\;b}}
\end{picture}
\begin{picture}(4,5)(-0.75,0)
\put(3,2.25){\makebox(0,0){\Huge-}}
\put(0,1.5){\Elline{2.5}{2}{r\;}{}}
\put(2,1.5){\Elline{2.5}{2}{}{\;s}} \put(0,2.5){\elstat{}{}{}{}}
\put(0,1.75){\PhotonNe{k\;}{}{}}
\put(0,0.5){\Elline{1}{0.5}{a\;}{}}
\put(2,0.5){\Elline{1}{0.5}{}{\;b}}
\end{picture}
\begin{picture}(4,4.5)(-1,0)
\put(0,2.3){\Elline{2.5}{2}{r\;}{}}
\put(0,2.3){\Elline{1}{0.5}{c\;}{}}
\put(2,2.3){\Elline{2.5}{0.5}{}{\;d}}
 \put(0,3.3){\PhotonNe{k\;}{}{}}
\put(0,0.5){\Elline{1}{0.5}{a\;}{}}
\put(2,0.5){\Elline{1}{0.5}{}{\;b}}
\put(-0.25,1.5){\Ebox{2.5}{0.8}}\put(1,1.9){\makebox(0,0){$V\eff$}}
\end{picture}
\\\begin{picture}(4,4.5)(-0.75,0)
\put(-2,2.25){\makebox(0,0){\large$\rho_{\G,ab}$\;=}}
\put(3.1,2.25){\makebox(0,0){\LARGE=}}
\put(0,1.5){\Elline{2.5}{2}{r\;}{}}
\put(2,1.5){\Elline{2.5}{2}{}{\;s}} \put(0,2.25){\Photon{}{}{}}
\put(0,0.5){\Elline{1}{0.5}{a\;}{}}
\put(2,0.5){\Elline{1}{0.5}{}{\;b}}
\end{picture}
\begin{picture}(4,4.5)(-0.75,0)
\put(3.1,2.25){\makebox(0,0){\Large+}}
\put(0,1.5){\Elline{2.5}{2}{r\;}{}}
\put(2,1.5){\Elline{2.5}{2}{}{\;s}}
\put(0,1.5){\PhotonNe{k\;}{}{}} \put(0,1.5){\photonNE{}{}{}}
\put(0,0.5){\Elline{1}{0.5}{a\;}{}}
\put(2,0.5){\Elline{1}{0.5}{}{\;b}}
\end{picture}
\begin{picture}(4,4.5)(-0.75,0)
\put(3.1,2.25){\makebox(0,0){\Huge-}}
\put(0,1.5){\Elline{2.5}{2}{r\;}{}}
\put(2,1.5){\Elline{2.5}{2}{}{\;s}} \put(0,2.75){\elstat{}{}{}}
\put(0,1.75){\Photon{}{}{}} \put(0,0.5){\Elline{1}{0.5}{a\;}{}}
\put(2,0.5){\Elline{1}{0.5}{}{\;b}}
\end{picture}
\begin{picture}(4,4.5)(-1,0)
\put(0,2.3){\Elline{2.1}{1.6}{r\;}{}}
\put(2,2.3){\Elline{2.1}{1.6}{}{\;s}}
\put(0,2.3){\Elline{1}{0.5}{c\;\;\;}{}}
\put(2,2.3){\Elline{1}{0.5}{}{\;\;\;d}}
 \put(0,3.3){\Photon{}{}{}}
\put(0,0.5){\Elline{1}{0.5}{a\;}{}}
\put(2,0.5){\Elline{1}{0.5}{}{\;b}}
\put(-0.25,1.5){\Ebox{2.5}{0.8}}\put(1,1.9){\makebox(0,0){$V\eff$}}
\end{picture}
\renewcommand{\normalsize}{\footnotesize}
\caption{Graphical representation the pair equations \eqref{BE3b},
solving the Bloch equation equations (\ref{BlochG}) and
\eqref{BlochGG}.}
\renewcommand{\normalsize}{\standard}
\label{Fig:BlochG}
\end{center}
\end{figure}
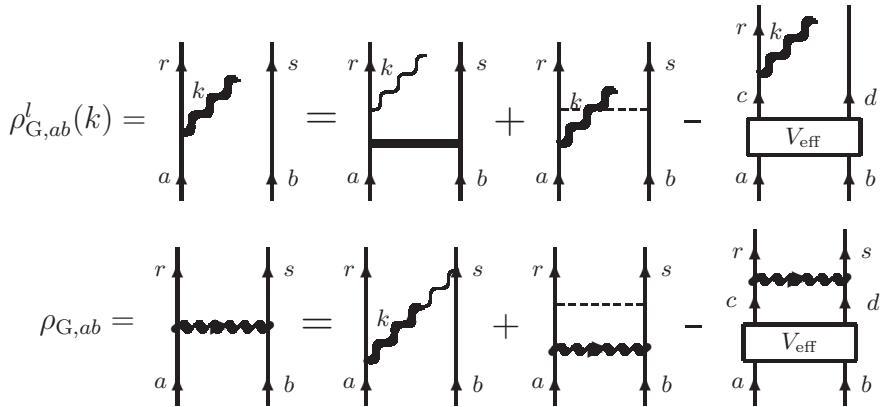

In order to treat the case where the instantaneous interactions
cross a retarded photon, we have to apply the former \it{between}
the two perturbations of the retarded interaction. We denote the
wave operator with an "uncontracted" retarded photon and an
arbitrary number of instantaneous interactions before and after
the retarded photon is created by $\Om^l_\G(k)$ and
$\Om^l_\sr(k)$, respectively, for the two components of the
retarded interaction. The components of these operators with no
model-space contributions are in analogy with previous cases
denoted by $\Ombar^l_\G(k)$ and $\Ombar^l_\sr(k)$, respectively.
We then have for the Gaunt interaction (and similarly in the
scalar-retardation case)
\begin{equation}
  \label{OmG}
  \Ombar_\G^l(k)P=\big(1+\Gk\VI+\Gk\VI\Gk\VI+\cdots\big)\Gk V_\G^l(k)
  \big(1+\GQ\VI+\GQ\VI\GQ\VI+\cdots\big)P.
\end{equation}
The rightmost bracket represents the wave operator $\Ombar_\mI$
\eqref{OmbarI1}, which leads to
\begin{equation}
  \label{OmG2}
  \Ombar_\G^l(k)P=\Gk V_\G^l(k)\Ombar_\mI P+\Gk\VI\Ombar_\G^l(k)P.
\end{equation}
Inserting this into the expression \eqref{116}, yields
\begin{equation}
  \label{BlochG0}
  Q\Om_\G^l(k)P=\bGQ\sum_{n=0}^\infty\pdn{n}{(V_\G^l(k)\Ombar_\mI)}\,(V\eff)^n
  +\bGQ\sum_{n=0}^\infty\pdn{n}{(\VI\Ombar_\G^l(k))}\,(V\eff)^n-\bGQ\Om_\G^l(k)V\eff,
\end{equation}
Since $\VI$ as well as $V_\G^l(k)$ are energy independent, we have
\begin{equation}
  \label{DerProd}
  \sum_{n=0}^\infty\pdn{n}{(V_\G^l(k)\Ombar_\mI)}\,(V\eff)^n=
  V_\G^l(k)\sum_{n=0}^\infty\pdn{n}{\Ombar_\mI}\,(V\eff)^n
  =V_\G^l(k)\,\Om_\mI P
\end{equation}
again using the expansion \eqref{Key} and the rule \eqref{Diffn}.
Treating the second sum similarly, leads to
\begin{equation}
  \label{BlochG1}
  Q\Om_\G^l(k)P=\bGQ V_\G^l(k)\,\Om_\mI P
  +\bGQ\VI\Om_\G^l(k)P-\bGQ\Om_\G^l(k)V\eff
\end{equation}
and to the Bloch equation
\begin{equation}
  \label{BlochG}
  \boxed{\big[\Om_\G^l(k),\bH_0\big]P
  =V_\G^l(k)\Om_\mI P+\VI\Om_\G^l(k)P-\Om_\G^l(k)V\eff}
\end{equation}

In the next step the photon of the function $\Om_\G^l(k)$ is being
absorbed by the other electron, followed by additional Coulomb
iterations. We denote the corresponding wave operator by $\Om_G$.
Omitting for the time being the MSC associated with the Coulomb
interactions, this leads to
\begin{equation}
  \label{OmGG}
  \Ombar_G P=\big(1+\GQ\VI+\GQ\VI\GQ\VI+\cdots\big)\,\GQ
  V_\G^l(k)\,\Om_\G^l(k)P,
\end{equation}
integrating over $k$ and summing over $l$ according to the
relation \eqref{fk}. The MSC are as before obtained by inserting
this relation into the formula \eqref{116}, which---using the same
argument as before---yields
\begin{equation}
  \label{BlochGG}
  \boxed{\big[\Om_\G,H_0\big]P
  =V_\G^l(k)\,\Om_\G^l(k)P+\VI \Om_\G P-\Om_\G V\eff}
\end{equation}

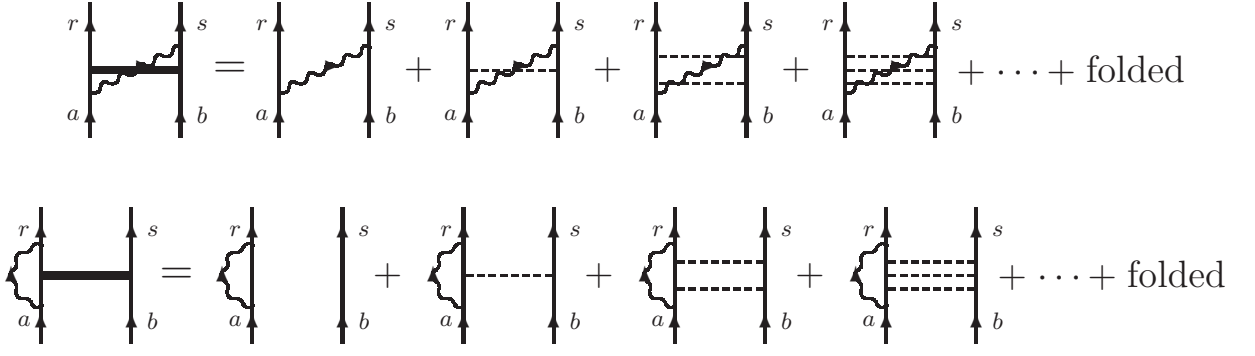
\begin{figure}
\begin{center}
\setlength{\unitlength}{0.6cm}
\begin{picture}(4,4)(-0.75,0)
\put(3.1,2){\makebox(0,0){\LARGE=}}
\put(0,1.5){\Elline{2}{1.5}{r\;}{}}
\put(2,1.5){\Elline{2}{1.5}{}{\;s}} \put(0,1.5){\photonENE{}{}{}}
{\linethickness{1mm}\put(0,2){\line(1,0){2}}}
\put(0,0.5){\Elline{1}{0.5}{a\;}{}}
\put(2,0.5){\Elline{1}{0.5}{}{\;b}}
\end{picture}
\begin{picture}(4,4)(-0.75,0)
\put(3.1,2){\makebox(0,0){\Large +}}
\put(0,1.5){\Elline{2}{1.5}{r\;}{}}
\put(2,1.5){\Elline{2}{1.5}{}{\;s}}
\put(0,0.5){\Elline{1}{0.5}{a\;}{}} \put(0,1.5){\photonENE{}{}{}}
\put(2,0.5){\Elline{1}{0.5}{}{\;b}}
\end{picture}
\begin{picture}(4,4)(-0.75,0)
\put(3.1,2){\makebox(0,0){\Large +}}
\put(0,1.5){\Elline{2}{1.5}{r\;}{}}
\put(2,1.5){\Elline{2}{1.5}{}{\;s}}
\put(0,0.5){\Elline{1}{0.5}{a\;}{}} \put(0,2){\elstat{}{}{}}
\put(0,1.5){\photonENE{}{}{}} \put(2,0.5){\Elline{1}{0.5}{}{\;b}}
\end{picture}
\begin{picture}(4,4)(-0.75,0)
\put(3.1,2){\makebox(0,0){\Large +}}
\put(0,1.5){\Elline{2}{1.5}{r\;}{}}
\put(2,1.5){\Elline{2}{1.5}{}{\;s}}
\put(0,0.5){\Elline{1}{0.5}{a\;}{}} \put(0,2.3){\elstat{}{}{}}
\put(0,1.7){\elstat{}{}{}} \put(0,1.5){\photonENE{}{}{}}
\put(2,0.5){\Elline{1}{0.5}{}{\;b}}
\end{picture}
\begin{picture}(4,4)(-0.75,0)
\put(5,2){\makebox(0,0){\Large +\;$\cdots$+\;folded}}
\put(0,1.5){\Elline{2}{1.5}{r\;}{}}
\put(2,1.5){\Elline{2}{1.5}{}{\;s}}
\put(0,0.5){\Elline{1}{0.5}{a\;}{}} \put(0,2.3){\elstat{}{}{}}
 \put(0,2){\elstat{}{}{}} \put(0,1.7){\elstat{}{}{}}
\put(0,1.5){\photonENE{}{}{}} \put(2,0.5){\Elline{1}{0.5}{}{\;b}}
\end{picture}
\\\begin{picture}(4,4)(-0.75,0)
\put(3,2){\makebox(0,0){\LARGE=}}
\put(0,1.5){\Elline{2}{1.5}{r\;}{}}
\put(2,1.5){\Elline{2}{1.5}{}{\;s}}
{\setlength{\unitlength}{0.4cm}\put(0,3){\ElSEL{}{}{}{}}}
{\linethickness{1mm}\put(0,2){\line(1,0){2}}}
\put(0,0.5){\Elline{1}{0.5}{a\;}{}}
\put(2,0.5){\Elline{1}{0.5}{}{\;b}}
\end{picture}
\begin{picture}(4.5,4)(-1.25,0)
\put(3,2){\makebox(0,0){\Large +}}
\put(0,1.5){\Elline{2}{1.5}{r\;}{}}
\put(2,1.5){\Elline{2}{1.5}{}{\;s}}
\put(0,0.5){\Elline{1}{0.5}{a\;}{}}
{\setlength{\unitlength}{0.4cm}\put(0,3){\ElSEL{}{}{}{}}}
\put(2,0.5){\Elline{1}{0.5}{}{\;b}}
\end{picture}
\begin{picture}(4.5,4.5)(-1.25,0)
\put(3,2){\makebox(0,0){\Large +}}
\put(0,1.5){\Elline{2}{1.5}{r\;}{}}
\put(2,1.5){\Elline{2}{1.5}{}{\;s}}
\put(0,0.5){\Elline{1}{0.5}{a\;}{}} \put(0,2){\elstat{}{}{}}
{\setlength{\unitlength}{0.4cm}\put(0,3){\ElSEL{}{}{}{}}}
\put(2,0.5){\Elline{1}{0.5}{}{\;b}}
\end{picture}
\begin{picture}(4.5,4)(-1.25,0)
\put(3,2){\makebox(0,0){\Large +}}
\put(0,1.5){\Elline{2}{1.5}{r\;}{}}
\put(2,1.5){\Elline{2}{1.5}{}{\;s}}
\put(0,0.5){\Elline{1}{0.5}{a\;}{}} \put(0,2.3){\elstat{}{}{}}
\put(0,1.7){\elstat{}{}{}}
{\setlength{\unitlength}{0.4cm}\put(0,3){\ElSEL{}{}{}{}}}
\put(2,0.5){\Elline{1}{0.5}{}{\;b}}
\end{picture}
\begin{picture}(4.5,4)(-1.25,0)
\put(5,2){\makebox(0,0){\Large +\;$\cdots$+\;folded}}
\put(0,1.5){\Elline{2}{1.5}{r\;}{}}
\put(2,1.5){\Elline{2}{1.5}{}{\;s}}
\put(0,0.5){\Elline{1}{0.5}{a\;}{}} \put(0,2.3){\elstat{}{}{}}
\put(0,1.7){\elstat{}{}{}}\put(0,2){\elstat{}{}{}}
{\setlength{\unitlength}{0.4cm}\put(0,3){\ElSEL{}{}{}{}}}
\put(2,0.5){\Elline{1}{0.5}{}{\;b}}
\end{picture}
\renewcommand{\normalsize}{\footnotesize}
\caption{Graphical representation of a single retarded photon with
crossing Coulomb interactions and without Coulomb interactions
before and after the retarded interaction.}
\renewcommand{\normalsize}{\standard}
\label{Fig:PhotCoul}
\end{center}
\end{figure}

\begin{figure}
\begin{center}
\setlength{\unitlength}{0.6cm}
\begin{picture}(4,4.5)(-0.5,0)
\put(-2,1.7){\makebox(0,0){\Large$\rho_{ab}^\con$\;=}}
\put(3.1,1.7){\makebox(0,0){\LARGE=}}
\put(0,1){\Elline{2.5}{2}{r\;}{}}
\put(2,1){\Elline{2.5}{2}{}{\;s}}
\put(0,0){\Elline{1}{0.5}{a\;}{}}
\put(2,0){\Elline{1}{0.5}{}{\;b}}
{\linethickness{1mm}\put(0,0.9){\line(1,0){2}}}
{\linethickness{1mm}\put(0,1.75){\line(1,0){2}}}
\put(0,1.2){\photonENE{}{}{}}
{\linethickness{1mm}\put(0,2.6){\line(1,0){2}}}
\end{picture}
\begin{picture}(5,4.5)(-0.5,0)
\put(3.6,1.7){\makebox(0,0){\LARGE+\,$\cdots$}}
\put(0,1){\Elline{2.5}{2}{r\;}{}}
\put(2,1){\Elline{2.5}{2}{}{\;s}}
\put(0,0){\Elline{1}{0.5}{a\;}{}}
\put(2,0){\Elline{1}{0.5}{}{\;b}}\put(0,1.2){\photonENE{}{}{}}
\end{picture}
\begin{picture}(4,5)(-0.5,0)
\put(3.6,1.7){\makebox(0,0){\LARGE+\,$\cdots$}}
\put(0,1){\Elline{3}{2.5}{r\;}{}}
\put(2,1){\Elline{3}{2.5}{}{\;s}}
\put(0,-0.5){\Elline{1.5}{0.5}{a\;}{}}
\put(2,-0.5){\Elline{1.5}{0.5}{}{\;b}} \put(0,0.5){\Melstat{3}}
\put(0,1.55){\Melstat{3}} \put(0,1.2){\photonENE{}{}{}}
\put(0,2.6){\Melstat{3}}
\end{picture}
\\\begin{picture}(4.5,4.5)(-0.5,-0.5)
\put(3.1,1.7){\makebox(0,0){\LARGE=}}
\put(0,1){\Elline{2.5}{2}{r\;}{}}
\put(2,1){\Elline{2.5}{2}{}{\;s}}
\put(0,0){\Elline{1}{0.5}{a\;}{}}
\put(2,0){\Elline{1}{0.5}{}{\;b}}
{\linethickness{1mm}\put(0,0.9){\line(1,0){2}}}
{\linethickness{1mm}\put(0,1.75){\line(1,0){2}}}
{\setlength{\unitlength}{0.35cm}\put(0,3){\ElSEL{}{}{}{}}}
{\linethickness{1mm}\put(0,2.6){\line(1,0){2}}}
\end{picture}
\begin{picture}(6,4.5)(-0.5,-0.5)
\put(3.6,1.7){\makebox(0,0){\LARGE+\,$\cdots$}}
\put(0,1){\Elline{2.5}{2}{r\;}{}}
\put(2,1){\Elline{2.5}{2}{}{\;s}}
\put(0,0){\Elline{1}{0.5}{a\;}{}}
\put(2,0){\Elline{1}{0.5}{}{\;b}}
{\setlength{\unitlength}{0.35cm}\put(0,3){\ElSEL{}{}{}{}}}
\end{picture}
\begin{picture}(4.5,5.5)(-0.5,-0.5)
\put(3.6,1.7){\makebox(0,0){\LARGE+\,$\cdots$}}
\put(0,1){\Elline{3}{2.5}{r\;}{}}
\put(2,1){\Elline{3}{2.5}{}{\;s}}
\put(0,-0.5){\Elline{1.5}{0.5}{a\;}{}}
\put(2,-0.5){\Elline{1.5}{0.5}{}{\;b}} \put(0,0.5){\Melstat{3}}
\put(0,1.55){\Melstat{3}}
{\setlength{\unitlength}{0.35cm}\put(0,3){\ElSEL{}{}{}{}}}
\put(0,2.6){\Melstat{3}}
\end{picture}
\renewcommand{\normalsize}{\footnotesize} \caption{Graphical
representation of a single retarded photon with crossing Coulomb
interactions and with Coulomb interactions before and after the
retarded interaction. The diagrams in upper line are obtained by
solving the pair equations \eqref{BE3b} and those in lower line by
analogous equation. }
\renewcommand{\normalsize}{\standard}
\label{Fig:PairPhotCoul}
\end{center}
\end{figure}
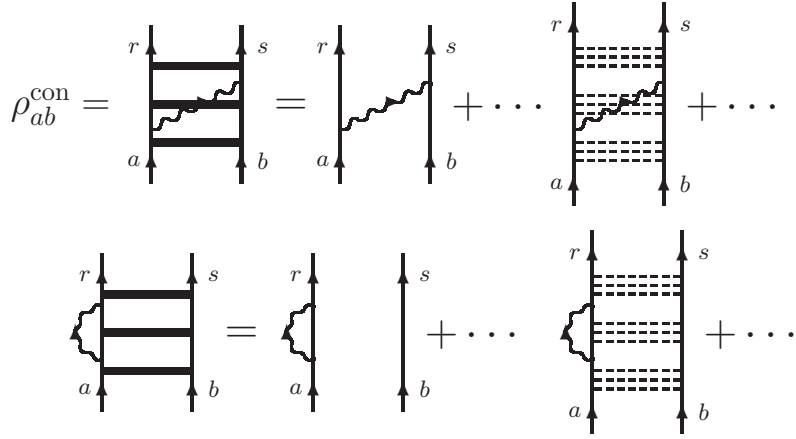

The framed equations above represent our main equations for
dealing with the combined retarded and unretarded interactions.
For a two-electron system they can be converted to the pair
equations
\begin{eqnarray}
  \label{BE3b}
  \big(\eps_{a}+\eps_{b}-h_0(1)-h_0(2)-k\big)
  \ket{\rho_{\G,ab}^l(k)}&=&\ip{rs}V^{l}_G(k)\ket{\rho_{ab}}
  +\ip{rs}\VI\ket{\rho_{\G,ab}^l(k)}
  -\ket{\rho_{\G,cd}^l(k)}\bra{cd}V\eff\ket{ab}\nn
  \big(\eps_{a}+\eps_{b}-h_0(1)-h_0(2)\big)\ket{\rho_{\G,ab}} &=&
  \ip{rs}V^{l}_G(k)\ket{\rho_{\G,ab}^l(k)}
  +\ip{rs}\VI\ket{\rho_{\G,ab}}
  -\ket{\rho_{cd}}\bra{cd}V\eff\ket{ab}.\Hsp
\end{eqnarray}
Note that the first of these equations is deduced from the
relation \eqref{BlochG} with the extended $\bH_0$, which leads to
the momentum $k$ on the left-hand side. These equations, which can
be solved using standard technique~\cite{SO89,SO89a}, are
illustrated in Fig. \ref{Fig:BlochG}. In Figs \ref{Fig:PhotCoul}
and \ref{Fig:PairPhotCoul} (upper line) we show more explicitly
the diagrams involved in the process just described---in the
former case without any Coulomb interactions before and after the
retarded interaction and in the latter case with such
interactions. When the photon is instead absorbed by the same
electron as it is emitted from, we get the corresponding
self-energy and vertex-correction effects---of course, after
appropriate renormalization--- as indicated in the bottom lines of
the same figures.

\subsection{Derivation of the general Bloch equations}

The Bloch equations derived above are valid only in the case of no
more than one uncontracted photon at each instance. In order to
derive the more general \it{Fock-space Bloch equation}, we can
proceed exactly as in the energy-independent case~\cite{Li74},
starting from the Schr\ödinger-like equation \eqref{SCHR}. We
first project this equation on the model space
\begin{equation}
  \label{Schr1}
  P\big(\bH_0+H'\big)\OM\bs{\Psi}_0^\alpha
  =\bH_0\bs{\Psi}_0^\alpha+P H'\OM\bs{\Psi}_0^\alpha
  =E^\alpha\bs{\Psi}_0^\alpha,
\end{equation}
using the fact that $P$ and $\bH_0$ commute, and then operate from
the left with $\bs{\Om}$
\begin{equation}
  \label{Schr2}
  \bs{\Om}\bH_0\bs{\Psi}_0^\alpha+\bs{\Om}P H'\OM\bs{\Psi}_0^\alpha
  =E^\alpha\bs{\Psi}^\alpha.
\end{equation}
Subtracting the original SE~\eqref{Schr}, then yields the
\it{Fock-space Bloch equation}
\begin{equation}
  \label{BE}
  \boxed{\big[\OM,\bH_0\big]P=\v\OM P-\OM V\eff\,;
  \qquad \bH\eff=P\bH_0 P+V\eff\,;
  \qquad V\eff=P\v\OM P}
\end{equation}
The solution can be expressed
\begin{equation}
  \label{BE2}
  \boxed{\Q\OM P=\bGQ\big(\v\OM-\OM V\eff\big)P}
\end{equation}
where $\bGQ$ is the generalized resolvent \eqref{GammaQ1}.

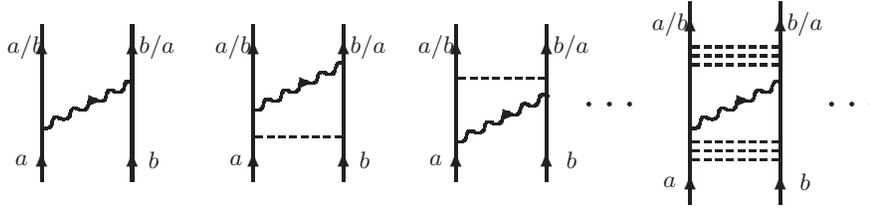
\begin{figure}
\begin{center}
\setlength{\unitlength}{0.6cm}
\begin{picture}(4.5,4.5)(-0.5,0)
\put(0,1){\Elline{2.5}{2}{a/b\;}{}}
\put(2,1){\Elline{2.5}{2}{}{\;\;b/a}}
\put(0,0){\Elline{1}{0.5}{a\;\;}{}}
\put(2,0){\Elline{1}{0.5}{}{\;b}}\put(0,1.2){\photonENE{}{}{}}
\end{picture}
\begin{picture}(4.5,4.5)(-0.5,0)
\put(0,1){\Elline{2.5}{2}{a/b\;\;}{}}
\put(2,1){\Elline{2.5}{2}{}{\;\;b/a}}
\put(0,0){\Elline{1}{0.5}{a\;}{}}
\put(2,0){\Elline{1}{0.5}{}{\;b}}\put(0,1.6){\photonENE{}{}{}}
\put(0,1){\elstat{}{}{}}
\end{picture}\begin{picture}(5,4.5)(-0.5,0)
\put(3.4,1.7){\makebox(0,0){\LARGE\,$\cdots$}}
\put(0,1){\Elline{2.5}{2}{a/b\;\;}{}}
\put(2,1){\Elline{2.5}{2}{}{\;\;b/a}}
\put(0,0){\Elline{1}{0.5}{a\;\;}{}}
\put(2,0){\Elline{1}{0.5}{}{\;\;b}}\put(0,0.9){\photonENE{}{}{}}
\put(0,2.3){\elstat{}{}{}}
\end{picture}
\begin{picture}(4.5,5)(-0.5,0)
\put(3.6,1.7){\makebox(0,0){\LARGE\,$\cdots$}}
\put(0,1){\Elline{3}{2.5}{a/b\;\;}{}}
\put(2,1){\Elline{3}{2.5}{}{\;\;b/a}}
\put(0,-0.5){\Elline{1.5}{0.5}{a\;\;}{}}
\put(2,-0.5){\Elline{1.5}{0.5}{}{\;\;b}} \put(0,0.5){\Melstat{3}}
 \put(0,1.2){\photonENE{}{}{}}
\put(0,2.6){\Melstat{3}}
\end{picture}
\renewcommand{\normalsize}{\footnotesize} \caption{Diagrams evaluated for the $1s2s\,^1S$
and $1s2s\,^3S$ states of heliumlike neon with the results given
in Table \ref{tab:Ne1s2s}.}
\renewcommand{\normalsize}{\standard}
\label{Fig:1s2sNe}
\end{center}
\end{figure}

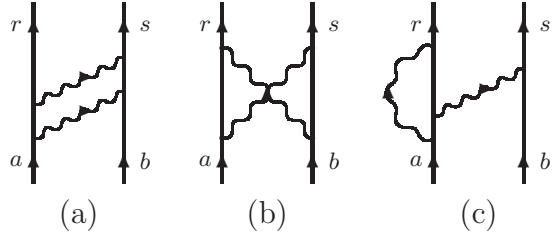
\begin{figure}
\begin{center}
\setlength{\unitlength}{0.6cm}
\begin{picture}(4,4.5)(-0.75,0)
\put(1,-0.2){\makebox(0,0){\large(a)}}
\put(0,1.5){\Elline{3}{2.5}{r\;}{}}
\put(2,1.5){\Elline{3}{2.5}{}{\;s}}
\put(0,0.5){\Elline{1}{0.5}{a\;}{}} \put(0,1.5){\photonENE{}{}{}}
 \put(0,2.25){\photonENE{}{}{}}
\put(2,0.5){\Elline{1}{0.5}{}{\;b}}
\end{picture}
\begin{picture}(4,4.5)(-0.75,0)
\put(1,-0.2){\makebox(0,0){\large(b)}}
\put(0,1.5){\Elline{3}{2.5}{r\;}{}}
\put(2,1.5){\Elline{3}{2.5}{}{\;s}}
\put(0,0.5){\Elline{1}{0.5}{a\;}{}}
\put(0,1.5){\Crossphotons{}{}{}{}{}{}}
\put(2,0.5){\Elline{1}{0.5}{}{\;b}}
\end{picture}
\begin{picture}(4.5,4.5)(-1.25,0)
\put(1,-0.2){\makebox(0,0){\large(c)}}
\put(0,1.5){\Elline{3}{2.5}{r\;}{}}
\put(2,1.5){\Elline{3}{2.5}{}{\;s}}
\put(0,0.5){\Elline{1}{0.5}{a\;}{}} \put(0,2){\photonENE{}{}{}}
{\setlength{\unitlength}{0.6cm}\put(0,2.5){\ElSEL{}{}{}{}}}
\put(2,0.5){\Elline{1}{0.5}{}{\;b}}
\end{picture}
\renewcommand{\normalsize}{\footnotesize}
\caption{Some irreducible diagrams, not included in the procedure
described here.}
\renewcommand{\normalsize}{\standard}
\label{Fig:Irred}
\end{center}
\end{figure}

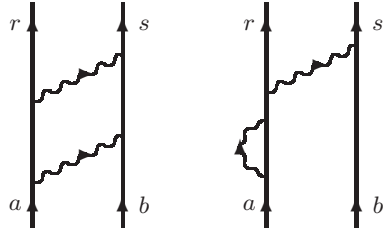
\begin{figure}[thb]
\begin{center}
{\setlength{\unitlength}{0.6cm}}
\begin{picture}(5,5.5)(-0.5,-0.5)
\put(0,2.8){\photonENE{}{}{}} \put(0,1){\Elline{4}{3.5}{r\;}{}}
\put(2,1){\Elline{4}{3.5}{}{\;s}}
\put(0,0){\Elline{1}{0.5}{a\;}{}}
\put(2,0){\Elline{1}{0.5}{}{\;b}} \put(0,1){\photonENE{}{}{}}
\end{picture}
\begin{picture}(5.5,5.5)(-0.5,-0.5)
\put(0,3){\photonENE{}{}{}} \put(0,1){\Elline{4}{3.5}{r\;}{}}
\put(2,1){\Elline{4}{3.5}{}{\;s}}
\put(0,0){\Elline{1}{0.5}{a\;}{}}
\put(2,0){\Elline{1}{0.5}{}{\;b}}
{\setlength{\unitlength}{0.35cm}\put(0,3){\ElSEL{}{}{}{}}}
\end{picture}
\renewcommand{\normalsize}{\footnotesize}
\caption{Examples of reducible diagrams, obtained by repeated use
of the procedure described.}
\renewcommand{\normalsize}{\standard}
\label{Fig:Red}
\end{center}
\end{figure}

We can expand the Fock-space wave operator and resolvent into
components acting in the subspace with no photons, with one photon
etc. as
\begin{eqnarray}
  \label{Omexp}
  \left\{\begin{array}{l}
  \OM=\Om+\Om^++\cdots\\
  \bGQ=\GQ+\GQ^+\cdots
  \end{array}\right.
\end{eqnarray}
The generalized Bloch equation \eqref{BE2} can then be separated
into
\begin{eqnarray}
  \label{BE3}
 \contract{1}{-7}{-1.7}
 \contract{1}{-8.8}{-0.6}
  \left\{\begin{array}{l}
  \Om P=P+\GQ\Big[\VI\Om+V_\rm{Ret}\Om^+-\Om V\eff \Big]P\\
  \Om^+ P=\GQ^+\Big[V_\rm{Ret}\Om+\VI\Om^++V_\rm{Ret}\Om^{++}-\Om^+V\eff\Big]P\\
  \Om^{++}P=\cdots\\
  \end{array}\right.
  \end{eqnarray}
The hook represents integration over the photon momentum $k$ and
summation over the angular momentum $l$, according to the
single-photon expression \eqref{fk}. These equations are valid
also in the case of multiple free photons at the same time.
Considering at most a single free photon, we see that they lead to
the Bloch equations \eqref{BlochG} and \eqref{BlochGG} derived
above.

\begin{table}
\caption{\mbox{Effects of one- and two-photon exchange for the
excited $1s2s\,^1S$ and $^3S$ states of heliumlike neon (in $\mu
H$).}}
\begin{tabular}{|c|c|c|c|}
\hline \hline & &$1s2s\,^1S$ & $1s2s\,^3S$  \\
\hline
One-photon &Gaunt & 2465.44 & 171,50 \\
&      Scalar ret. & 171,58 & -171,58 \\\hline
Two-photon &Coul.-Gaunt & -794.8 & -51.8 \\
&         Coul.-Scal.ret. & 22.5 & 42.5\\\hline
One + two-photon &Gaunt & 1670.7 & 119.7 \\
&         Scal.ret. & 194.2 & -129.1\\\hline
One photon correlated&Gaunt & 1752.0 & 124.6 \\
&         Scal.ret. & 183.9 & -132.2\\
 \hline \hline
\end{tabular}
\label{tab:Ne1s2s}
\end{table}

\section{Numerical procedure and results} The pair equations
considered here can be solved numerically with essentially the
same technique as developed by Salomonson and \Öster for
(relativistic) many-body calculations~\cite{SO89,SO89a} and used
in our previous works~\cite{LAS01,ASL02}. The radial integrations
are performed with an exponential grid with 70-150 grid points and
the $k$ integration with 100-150 points using Gaussian quadrature.
For excited states poles appear in the $k$ integration, which
require special attention (for details, see ref.~\cite{ASL02}).
The numerical calculations are quite time consuming in the present
case, since separate pair functions have to be evaluated for each
value of the photon momentum. On the other hand, the procedure is
particularly well suite for parallel computing, and we hope that
the procedure can be speeded up considerably, when our routines
are better optimized.

Here, we shall only give some illustrative examples of our
numerical results---more results will be published
separately~\cite{HSL06}. In Table \ref{tab:Ne1s2s} we show the
effect of one retarded photon with one and infinitely many
non-crossing Coulomb interactions for the $1s2s\,^1S$ and $^3S$
states of heliumlike neon. This corresponds to the diagrams shown
in Fig. \ref{Fig:1s2sNe}, excluding virtual pairs (NVP). This
represents the first numerical evaluation of effects beyond
two-photon exchange, involving a retarded interaction. The
two-photon effects have been compared with the corresponding
$S$-matrix results~\cite{ASL02} and are found to agree to 3-4
digits, which represents the numerical accuracy of the present
calculations. The effect of correlation beyond second order is in
this case found to be about five percent of the total
contribution, which is one order of magnitude larger than the
effect of the retarded (reducible and irreducible) two-photon
interaction~\cite{ASL02}. The effect of virtual pairs is of the
same order as that of two retarded interactions. This indicates
that for heliumlike neon the procedure described here with a
single retarded photon together with Coulomb interactions
represents about 99 \% of the non-radiative effects not included
in a standard many-body perturbation treatment with only
instantaneous Coulomb interactions. For lighter elements the
importance of a single retarded photon is even more pronounced.

\section{Summary and Outlook}
We have in previous articles described a new technique for QED
calculations that we refer to as the covariant evolution-operator
(CEO) method ~\cite{Li98,LAS01,LSA04,LSH05}. This method has the
great advantage compared to other QED techniques that it has a
structure very akin to that of standard many-body perturbation
theory, which opens up the possibility for a merger of the two
techniques. In two previous papers~\cite{LSA04,LSH05} we have
indicated how such a merger can be accomplished, and this is
developed further in the present paper, and some numerical results
are presented. Combined QED and correlation effects, which can be
treated only in a very limited fashion by standard techniques, are
of particular importance for light and medium-heavy elements. The
CEO method also has the advantage compared to the standard
$S$-matrix technique that it can be applied to the case of
quasidegeneracy (a property it shares with the two-times
Green's-function technique of Shabaev \it{et al.}~\cite{Shab02}).

The procedure we have developed represents the exchange of a
single retarded photon and an arbitrary number of instantaneous
Coulomb interactions between the electrons, crossing and
non-crossing. So far we have been working with positive-energy
intermediate states---no-virtual-pair (NVP) approximation---but
single and double virtual pairs can be included in the procedure.
The procedure can also be used---with proper
renormalizations---for radiative effects (self energy and vertex
corrections) with a single retarded photon (see Figs
\ref{Fig:PairPhotCoul} and \ref{Fig:PhotCoul}). In principle, the
procedure can be used also for \it{irreducible} multi-photon
effects, where retarded interactions overlap in time, like those
indicated in Fig. \ref{Fig:Irred}, by treating more than one
uncontracted photon at a particular time. At present, however,
this is beyond reach with the computers we have available. On the
other hand, \it{reducible} multi-photon effects, where the
interactions are separated in time, as illustrated in Fig.
\ref{Fig:Red}, can be included by repeated use of the procedure
described.

It has been demonstrated that one retarded photon with Coulomb
interaction represents by far the dominating part of the
non-radiative multi-photon exchange for light and medium-heavy
elements beyond the standard Coulomb correlation (of the order of
99 \% for heliumlike neon), and the situation can be expected to
be similar for the radiative part. The small effects due to two
irreducible retarded photons without Coulomb interactions can be
evaluated with standard QED methods, and higher-order effects can
with good accuracy be estimated by means of analytical
approximations. Therefore, it is our belief that the method
presented here, when the routines are fully developed, should be
able to produce accurate results for energy separations, such as
the fine-structure separations, for light and medium-heavy
elements, hopefully down to neutral helium.

\section*{Acknowledgements}
The authors wish to express their thanks to Bj\örn {\AA}s\'en for
valuable discussions. The work has been supported by the Swedish
Research Council.

\appendix
\section{Coulomb-gauge interaction}
\newcommand{\bC}{\boldsymbol{C}}
\newcommand{\g}{_{\mathrm{G}}}
\label{App:Coul} The terms in the Coulomb-gauge expression
\eqref{VExplC} can be separated into a product of two
single-particle potentials by the spherical wave expansion, using
the relation
\begin{eqnarray}
  \label{eq:Sph_Wave}
  \frac{\sin(kr_{12})}{r_{12}}=k\sum^{\infty}_{l=0}(2l+1)
  j_l(kr_1)j_l(kr_2)\bC^{l}(1) \bsdot \bC^{l}(2),
\end{eqnarray}
where $\bC^{l}$ is a spherical tensor, associated with the
spherical harmonics $Y^l$. The Gaunt term then becomes
\begin{eqnarray}
  \label{eq:Gaunt_Sep}
  -k\sum^{\infty}_{l=0}(2l+1)V^{l}\g(kr_1)\bsdot V^{l}\g(kr_2),
\end{eqnarray}
where $V^{l}\g(kr_i)$ is
\begin{eqnarray}
  \label{eq:G_Single_Pot}
  V^l\g(kr)=\balpha j_l(kr) \bC^{l}.
\end{eqnarray}
For the scalar-retardation term we use the relation~\cite[Sect.
5.7]{Edm57}, ~\cite[Part II, App. A]{LR74}
\begin{eqnarray}
  \label{eq:CaseStudies_1}\nonumber
  \nabla[f(r)C^{l}_{m}]&=&\frac{1}{2l+1}\Big[ -\sqrt{(l+1)(2l+3)}
  (\frac{d}{dr}-\frac{l}{r})f(r)\bC^{l,l+1}_{m}\\
  &&+\sqrt{l(2l-1)}(\frac{d}{dr}+\frac{l+1}{r})f(r)\bC^{l,l-1}_{m}\Big],
\end{eqnarray}
where $\bC^{l,l\pm1}_{m}$ is a vector, associated with the vector
spherical harmonics $\bs{Y}^{l,l\pm1}_{m}$, and the relation
\begin{eqnarray}
  \label{eq:CaseStudies_2}
  \balpha \bsdot \bC^{l,k}_{m} =
  \big\{ \balpha\; \bC^l\big\}^k_m,
\end{eqnarray}
where the left-hand side is a scalar product and the right-hand
side a tensor product. Together with~\cite{Arf84}
\begin{eqnarray}
  \label{eq:Bessel_Der}
  (\frac{d}{dr}-\frac{l}{r})j_l(kr)=-kj_{l+1}(kr)\\
  (\frac{d}{dr}+\frac{l+1}{r})j_l(kr)=kj_{l-1}(kr)
\end{eqnarray}
the final result of the scalar retardation term becomes
\begin{eqnarray}
  \label{eq:SR_sep}
  \sum^{\infty}_{l=0}\frac{k}{2l+1}V^{l}_\sr(kr_1)\bsdot
  V^{l}_\sr(kr_2),
\end{eqnarray}
where expression for the single-particle potentials for the scalar
retardation is written as
\begin{eqnarray}
  \label{eq:SR_Single_Pot}\nonumber
  V^{l}_\sr(kr)&=&
  \Big[\sqrt{(l+1)(2l+3)} j_{l+1}(kr)
  \big\{ \balpha\; \bC^{l+1} \big\}^{l}\\
  && +\sqrt{l(2l-1)} j_{l-1}(kr)
  \big\{ \balpha\; \bC^{l-1} \big\}^{l}\Big].
\end{eqnarray}
The function, $f(k)$, in the Coulomb gauge expression
\eqref{VExplC} then becomes
\begin{eqnarray}
  \label{eq:Coulomb_f_Single}
  f(k)=\frac{e^{2}k}{4\pi^2}\sum^{\infty}_{l=0}
  \Big[-(2l+1)V^{l}\g(kr_1)\bsdot V^{l}\g(kr_2)+
  \frac{1}{2l+1}V^{l}_\sr(kr_1)\bsdot V^{l}_\sr(kr_2)\Big].
\end{eqnarray}

\section{Rules for differentiation}\label{sec:Diff}
The difference ratios we use in the formalism presented here are
of a special kind and give rise to special handling rules (see
also ref.~\cite{LSH05}, App. E).

If $A(\E)$ is an operator function of the (energy) parameter $\E$,
then we define the first-order difference ratio
   \[\pd{A(\E)}=\pd{_{\E\E'}A(\E)}=\frac{A(\E)-A(\E')}{\E-\E'}.\]
   Then
\begin{equation}
  \label{Diff1}
  \pd{}[A(\E)B(\E)]
    =\pd{_{\E\E'}A(\E)}\,B(\E')+A(\E)\,\pd{_{\E\E'}B(\E)}.
\end{equation}
The second difference ratio is defined as
  \[\pdn{2}{A(\E)}=\pdn{2}{_{\E\E'\E''}A(\E)}
  =\pd{_{\E\E'}}\pd{_{\E\E''}A(\E)}\]
and generally
  \[\partdeltan{n}{_{\E\E'\cdots\E^n}}{\E}
  =\partdelta{_{\E\E'}}{\E}\partdelta{_{\E\E''}}{\E}\cdots
  \partdelta{_{\E\E^n}}{\E}.\]
It then follows that
\begin{eqnarray}
  \label{Diff2a}
  &&\pdn{2}{}[A(\E)B(\E)]=\pdn{2}{_{\E\E'\E''}}{}[A(\E)B(\E)]
  =\pd{_{\E\E'}}\pd{_{\E\E''}}{}[A(\E)B(\E)]\nn
  &&=\pd{_{\E\E'}}{}\Big[\pd{_{\E\E''}A(\E)}\,B(\E'')+A(\E)\,\pd{_{\E\E''}B(\E)}\Big]\nn
  &&=\pdn{2}{A_{\E\E'\E''}(\E)}B(\E'')+\partdelta{_{\E\E'}A(\E)}{\E}\,
  \partdelta{_{\E\E''}B(\E)}{\E}
  +A(\E)\partdeltan{2}{_{\E\E'\E''}B(\E)}{\E}.
\end{eqnarray}
It should be noted that the operator $B(\E'')$ is unaffected by
the differentiation $\delta_{\E\E'}$. With simplified notations we
then have
\begin{equation}
  \label{Diff2}
  \pdn{2}{(AB)}=\pdn{2}{A}\,B+\pd{A}\,\pd{B}+A\,\pdn{2}{B},
\end{equation}
which can be generalized to
\begin{equation}
  \label{Diffn}
  \boxed{\pdn{n}{(AB)}=\sum_{m=0}^n\pdn{m}{A}\;\pdn{{(n-m)}}{B}}
\end{equation}

In the case of complete degeneracy we have in first order
\begin{equation}
  \label{Degen1}
  \lim_{\E'\rarr\E}\pd{_{\E\E'}A(\E)}=\partder{A(\E)}{\E},
\end{equation}
while in second order we have
\begin{equation}
  \label{Degen2}
  \lim_{\E',\E''\rarr\E}\pdn{2}{_{\E\E'\E''}A(\E)}
  =\half\partdern{2}{A(\E)}{\E}
\end{equation}
and generally
\begin{equation}
  \label{Degenn}
  \boxed{\lim_{\E',\E''\cdots\rarr\E}\pdn{n}{_{\E\E'\E''}A(\E)}
  =\frac{1}{n!}\partdern{n}{A(\E)}{\E}}
\end{equation}

 If we have a product of
effective interactions, then they are separated by model-space
states, which generally have different energies,
  \[\cdots P''V\eff P'V\eff P,\]
where $P$ corresponds to the energy $\E$, $P'$ to $\E'$  etc. This
appears in multiple folded terms, and the effective interactions
have the corresponding energy parameter,
  \[\cdots P''V\eff(\E') P'V\eff(\E) P.\]
If we now form the difference ratio of this product
  \[\pd{}\big[\cdots P''V\eff(\E') P'V\eff(\E) P\big],\]
implying that the parameter $\E$ is changed, only the last factor
is affected, and
  \[\pd{}\big[\cdots P''V\eff(\E') P'V\eff(\E) P\big]
  =\cdots P''V\eff(\E')P'\;\pd{V\eff(\E)}P\]
or generally
\begin{equation}
  \label{DiffVeff}
  \boxed{\pd{(V\eff)^n}=(V\eff)^{(n-1)}\,\pd{V\eff}}
\end{equation}

\bibliographystyle{c:/Sty/prsty}


\begin{thebibliography}{37}
\expandafter\ifx\csname natexlab\endcsname\relax\def\natexlab#1{#1}\fi
\expandafter\ifx\csname bibnamefont\endcsname\relax
  \def\bibnamefont#1{#1}\fi
\expandafter\ifx\csname bibfnamefont\endcsname\relax
  \def\bibfnamefont#1{#1}\fi
\expandafter\ifx\csname citenamefont\endcsname\relax
  \def\citenamefont#1{#1}\fi
\expandafter\ifx\csname url\endcsname\relax
  \def\url#1{\texttt{#1}}\fi
\expandafter\ifx\csname urlprefix\endcsname\relax\def\urlprefix{URL }\fi
\providecommand{\bibinfo}[2]{#2}
\providecommand{\eprint}[2][]{\url{#2}}

\bibitem[{\citenamefont{Lindgren and Morrison}(1986)}]{LM86}
\bibinfo{author}{\bibfnamefont{I.}~\bibnamefont{Lindgren}} \bibnamefont{and}
  \bibinfo{author}{\bibfnamefont{J.}~\bibnamefont{Morrison}},
  \emph{\bibinfo{title}{\textit{Atomic Many-Body Theory}}}
  (\bibinfo{publisher}{Second edition, Springer-Verlag},
  \bibinfo{address}{Berlin}, \bibinfo{year}{1986}).

\bibitem[{\citenamefont{Lindgren}(1998)}]{Li98}
\bibinfo{author}{\bibfnamefont{I.}~\bibnamefont{Lindgren}},
  \bibinfo{journal}{Mol. Phys.} \textbf{\bibinfo{volume}{94}},
  \bibinfo{pages}{19} (\bibinfo{year}{1998}).

\bibitem[{\citenamefont{Lindgren et~al.}(2001)\citenamefont{Lindgren,
  {\AA}s\'{e}n, Salomonson, and M{\aa}rtensson-Pendrill}}]{LAS01}
\bibinfo{author}{\bibfnamefont{I.}~\bibnamefont{Lindgren}},
  \bibinfo{author}{\bibfnamefont{B.}~\bibnamefont{{\AA}s\'{e}n}},
  \bibinfo{author}{\bibfnamefont{S.}~\bibnamefont{Salomonson}},
  \bibnamefont{and} \bibinfo{author}{\bibfnamefont{A.-M.}
  \bibnamefont{M{\aa}rtensson-Pendrill}}, \bibinfo{journal}{Phys. Rev. A}
  \textbf{\bibinfo{volume}{64}}, \bibinfo{pages}{062505}
  (\bibinfo{year}{2001}).

\bibitem[{\citenamefont{Lindgren et~al.}(2004)\citenamefont{Lindgren,
  Salomonson, and {\AA}s\'{e}n}}]{LSA04}
\bibinfo{author}{\bibfnamefont{I.}~\bibnamefont{Lindgren}},
  \bibinfo{author}{\bibfnamefont{S.}~\bibnamefont{Salomonson}},
  \bibnamefont{and}
  \bibinfo{author}{\bibfnamefont{B.}~\bibnamefont{{\AA}s\'{e}n}},
  \bibinfo{journal}{Physics Reports} \textbf{\bibinfo{volume}{389}},
  \bibinfo{pages}{161} (\bibinfo{year}{2004}).

\bibitem[{\citenamefont{Lindgren et~al.}(2005)\citenamefont{Lindgren,
  Salomonson, and Hedendahl}}]{LSH05}
\bibinfo{author}{\bibfnamefont{I.}~\bibnamefont{Lindgren}},
  \bibinfo{author}{\bibfnamefont{S.}~\bibnamefont{Salomonson}},
  \bibnamefont{and}
  \bibinfo{author}{\bibfnamefont{D.}~\bibnamefont{Hedendahl}},
  \bibinfo{journal}{Can. J. Phys.} \textbf{\bibinfo{volume}{83}},
  \bibinfo{pages}{183} (\bibinfo{year}{2005}).

\bibitem[{\citenamefont{Mohr et~al.}(1998)\citenamefont{Mohr, Plunien, and
  Soff}}]{MPS98}
\bibinfo{author}{\bibfnamefont{P.~J.} \bibnamefont{Mohr}},
  \bibinfo{author}{\bibfnamefont{G.}~\bibnamefont{Plunien}}, \bibnamefont{and}
  \bibinfo{author}{\bibfnamefont{G.}~\bibnamefont{Soff}},
  \bibinfo{journal}{Physics Reports} \textbf{\bibinfo{volume}{293}},
  \bibinfo{pages}{227} (\bibinfo{year}{1998}).

\bibitem[{\citenamefont{Mohr and Sapirstein}(2000)}]{MS00}
\bibinfo{author}{\bibfnamefont{P.~J.} \bibnamefont{Mohr}} \bibnamefont{and}
  \bibinfo{author}{\bibfnamefont{J.}~\bibnamefont{Sapirstein}},
  \bibinfo{journal}{Phys. Rev. A} \textbf{\bibinfo{volume}{62}},
  \bibinfo{pages}{052501.1} (\bibinfo{year}{2000}).

\bibitem[{\citenamefont{Shabaev}(2002)}]{Shab02}
\bibinfo{author}{\bibfnamefont{V.~M.} \bibnamefont{Shabaev}},
  \bibinfo{journal}{Physics Reports} \textbf{\bibinfo{volume}{356}},
  \bibinfo{pages}{119} (\bibinfo{year}{2002}).

\bibitem[{\citenamefont{Artemyev et~al.}(2005)\citenamefont{Artemyev, Shabaev,
  Yerokhin, Plunien, and Soff}}]{AShab05}
\bibinfo{author}{\bibfnamefont{A.~N.} \bibnamefont{Artemyev}},
  \bibinfo{author}{\bibfnamefont{V.~M.} \bibnamefont{Shabaev}},
  \bibinfo{author}{\bibfnamefont{V.~A.} \bibnamefont{Yerokhin}},
  \bibinfo{author}{\bibfnamefont{G.}~\bibnamefont{Plunien}}, \bibnamefont{and}
  \bibinfo{author}{\bibfnamefont{G.}~\bibnamefont{Soff}},
  \bibinfo{journal}{Phys. Rev. A} \textbf{\bibinfo{volume}{71}},
  \bibinfo{pages}{062104} (\bibinfo{year}{2005}).

\bibitem[{\citenamefont{Drake}(1979)}]{Dr79}
\bibinfo{author}{\bibfnamefont{G.~W.~F.} \bibnamefont{Drake}},
  \bibinfo{journal}{Phys. Rev. A} \textbf{\bibinfo{volume}{19}},
  \bibinfo{pages}{1387} (\bibinfo{year}{1979}).

\bibitem[{\citenamefont{Drake}(1988)}]{Dr88}
\bibinfo{author}{\bibfnamefont{G.~W.~F.} \bibnamefont{Drake}},
  \bibinfo{journal}{Can. J. Phys.} \textbf{\bibinfo{volume}{66}},
  \bibinfo{pages}{586} (\bibinfo{year}{1988}).

\bibitem[{\citenamefont{Drake}(2002)}]{Drake02}
\bibinfo{author}{\bibfnamefont{G.~W.~F.} \bibnamefont{Drake}},
  \bibinfo{journal}{Can. J. Phys.} \textbf{\bibinfo{volume}{80}},
  \bibinfo{pages}{1195} (\bibinfo{year}{2002}).

\bibitem[{\citenamefont{Pachucki}(2002)}]{Pach02}
\bibinfo{author}{\bibfnamefont{K.}~\bibnamefont{Pachucki}},
  \bibinfo{journal}{J. Phys. B} \textbf{\bibinfo{volume}{35}},
  \bibinfo{pages}{13087} (\bibinfo{year}{2002}).

\bibitem[{\citenamefont{Pachucki and Sapirstein}(2002)}]{PSap02}
\bibinfo{author}{\bibfnamefont{K.}~\bibnamefont{Pachucki}} \bibnamefont{and}
  \bibinfo{author}{\bibfnamefont{J.}~\bibnamefont{Sapirstein}},
  \bibinfo{journal}{J. Phys. B} \textbf{\bibinfo{volume}{35}},
  \bibinfo{pages}{1783} (\bibinfo{year}{2002}).

\bibitem[{\citenamefont{Sucher}(1957)}]{Su57a}
\bibinfo{author}{\bibfnamefont{J.}~\bibnamefont{Sucher}},
  \bibinfo{journal}{Phys. Rev.} \textbf{\bibinfo{volume}{109}},
  \bibinfo{pages}{1010} (\bibinfo{year}{1957}).

\bibitem[{\citenamefont{Douglas and Kroll}(1974)}]{DK74}
\bibinfo{author}{\bibfnamefont{M.~H.} \bibnamefont{Douglas}} \bibnamefont{and}
  \bibinfo{author}{\bibfnamefont{N.~M.} \bibnamefont{Kroll}},
  \bibinfo{journal}{Ann. Phys. (N.Y.)} \textbf{\bibinfo{volume}{82}},
  \bibinfo{pages}{89} (\bibinfo{year}{1974}).

\bibitem[{\citenamefont{Fritzsche et~al.}(2005)\citenamefont{Fritzsche,
  Indelicato, and St\"ohlker}}]{FIS05}
\bibinfo{author}{\bibfnamefont{S.}~\bibnamefont{Fritzsche}},
  \bibinfo{author}{\bibfnamefont{P.}~\bibnamefont{Indelicato}},
  \bibnamefont{and}
  \bibinfo{author}{\bibfnamefont{T.}~\bibnamefont{St\"ohlker}},
  \bibinfo{journal}{J. Phys. B} \textbf{\bibinfo{volume}{38}},
  \bibinfo{pages}{S707} (\bibinfo{year}{2005}).

\bibitem[{\citenamefont{Hedendahl et~al.}(to be
  published)\citenamefont{Hedendahl, Salomonson, and Lindgren}}]{HSL06}
\bibinfo{author}{\bibfnamefont{D.}~\bibnamefont{Hedendahl}},
  \bibinfo{author}{\bibfnamefont{S.}~\bibnamefont{Salomonson}},
  \bibnamefont{and} \bibinfo{author}{\bibfnamefont{I.}~\bibnamefont{Lindgren}}
  (\bibinfo{year}{to be published}).

\bibitem[{\citenamefont{Bloch}(1958{\natexlab{a}})}]{Bl58a}
\bibinfo{author}{\bibfnamefont{C.}~\bibnamefont{Bloch}},
  \bibinfo{journal}{Nucl. Phys.} \textbf{\bibinfo{volume}{6}},
  \bibinfo{pages}{329} (\bibinfo{year}{1958}{\natexlab{a}}).

\bibitem[{\citenamefont{Bloch}(1958{\natexlab{b}})}]{Bl58b}
\bibinfo{author}{\bibfnamefont{C.}~\bibnamefont{Bloch}},
  \bibinfo{journal}{Nucl. Phys.} \textbf{\bibinfo{volume}{7}},
  \bibinfo{pages}{451} (\bibinfo{year}{1958}{\natexlab{b}}).

\bibitem[{\citenamefont{Lindgren}(1974)}]{Li74}
\bibinfo{author}{\bibfnamefont{I.}~\bibnamefont{Lindgren}},
  \bibinfo{journal}{J. Phys. B} \textbf{\bibinfo{volume}{7}},
  \bibinfo{pages}{2441} (\bibinfo{year}{1974}).

\bibitem[{\citenamefont{Lindgren}(1978)}]{Li78}
\bibinfo{author}{\bibfnamefont{I.}~\bibnamefont{Lindgren}},
  \bibinfo{journal}{Int. J. Quantum Chem.} \textbf{\bibinfo{volume}{S12}},
  \bibinfo{pages}{33} (\bibinfo{year}{1978}).

\bibitem[{\citenamefont{Fetter and Walecka}(1971)}]{FW71}
\bibinfo{author}{\bibfnamefont{A.~L.} \bibnamefont{Fetter}} \bibnamefont{and}
  \bibinfo{author}{\bibfnamefont{J.~D.} \bibnamefont{Walecka}},
  \emph{\bibinfo{title}{\textit{The Quantum Mechanics of Many-Body Systems}}}
  (\bibinfo{publisher}{McGraw-Hill}, \bibinfo{address}{N.Y.},
  \bibinfo{year}{1971}).

\bibitem[{\citenamefont{Gell-Mann and Low}(1951)}]{GML51}
\bibinfo{author}{\bibfnamefont{M.}~\bibnamefont{Gell-Mann}} \bibnamefont{and}
  \bibinfo{author}{\bibfnamefont{F.}~\bibnamefont{Low}},
  \bibinfo{journal}{Phys. Rev.} \textbf{\bibinfo{volume}{84}},
  \bibinfo{pages}{350} (\bibinfo{year}{1951}).

\bibitem[{\citenamefont{Goldstone}(1957)}]{Go57}
\bibinfo{author}{\bibfnamefont{J.}~\bibnamefont{Goldstone}},
  \bibinfo{journal}{Proc. R. Soc. London, Ser. A}
  \textbf{\bibinfo{volume}{239}}, \bibinfo{pages}{267} (\bibinfo{year}{1957}).

\bibitem[{\citenamefont{Schweber}(1961)}]{Sch61}
\bibinfo{author}{\bibfnamefont{S.~S.} \bibnamefont{Schweber}},
  \emph{\bibinfo{title}{\textit{An Introduction to Relativistic Quantum Field
  Theory}}} (\bibinfo{publisher}{Harper and Row}, \bibinfo{address}{N.Y.},
  \bibinfo{year}{1961}).

\bibitem[{\citenamefont{M{\aa}rtensson}(1980)}]{Ma79}
\bibinfo{author}{\bibfnamefont{A.-M.} \bibnamefont{M{\aa}rtensson}},
  \bibinfo{journal}{J. Phys. B} \textbf{\bibinfo{volume}{12}},
  \bibinfo{pages}{3995} (\bibinfo{year}{1980}).

\bibitem[{\citenamefont{Lindgren}(1985)}]{Li85}
\bibinfo{author}{\bibfnamefont{I.}~\bibnamefont{Lindgren}},
  \bibinfo{journal}{Phys. Rev. A} \textbf{\bibinfo{volume}{31}},
  \bibinfo{pages}{1273} (\bibinfo{year}{1985}).

\bibitem[{\citenamefont{Lindroth}(1988)}]{ELi88}
\bibinfo{author}{\bibfnamefont{E.}~\bibnamefont{Lindroth}},
  \bibinfo{journal}{Phys. Rev. A} \textbf{\bibinfo{volume}{37}},
  \bibinfo{pages}{316} (\bibinfo{year}{1988}).

\bibitem[{\citenamefont{Salomonson and {\"O}ster}(1989{\natexlab{a}})}]{SO90}
\bibinfo{author}{\bibfnamefont{S.}~\bibnamefont{Salomonson}} \bibnamefont{and}
  \bibinfo{author}{\bibfnamefont{P.}~\bibnamefont{{\"O}ster}},
  \bibinfo{journal}{Phys. Rev. A} \textbf{\bibinfo{volume}{41}},
  \bibinfo{pages}{4670} (\bibinfo{year}{1989}{\natexlab{a}}).

\bibitem[{\citenamefont{Salomonson and Ynnerman}(1991)}]{SY91}
\bibinfo{author}{\bibfnamefont{S.}~\bibnamefont{Salomonson}} \bibnamefont{and}
  \bibinfo{author}{\bibfnamefont{A.}~\bibnamefont{Ynnerman}},
  \bibinfo{journal}{Phys. Rev. A} \textbf{\bibinfo{volume}{43}},
  \bibinfo{pages}{88} (\bibinfo{year}{1991}).

\bibitem[{\citenamefont{Salomonson and {\"O}ster}(1989{\natexlab{b}})}]{SO89}
\bibinfo{author}{\bibfnamefont{S.}~\bibnamefont{Salomonson}} \bibnamefont{and}
  \bibinfo{author}{\bibfnamefont{P.}~\bibnamefont{{\"O}ster}},
  \bibinfo{journal}{Phys. Rev. A} \textbf{\bibinfo{volume}{40}},
  \bibinfo{pages}{5548} (\bibinfo{year}{1989}{\natexlab{b}}).

\bibitem[{\citenamefont{Salomonson and {\"O}ster}(1989{\natexlab{c}})}]{SO89a}
\bibinfo{author}{\bibfnamefont{S.}~\bibnamefont{Salomonson}} \bibnamefont{and}
  \bibinfo{author}{\bibfnamefont{P.}~\bibnamefont{{\"O}ster}},
  \bibinfo{journal}{Phys. Rev. A} \textbf{\bibinfo{volume}{40}},
  \bibinfo{pages}{5559} (\bibinfo{year}{1989}{\natexlab{c}}).

\bibitem[{\citenamefont{{\AA}s\'{e}n et~al.}(2002)\citenamefont{{\AA}s\'{e}n,
  Salomonson, and Lindgren}}]{ASL02}
\bibinfo{author}{\bibfnamefont{B.}~\bibnamefont{{\AA}s\'{e}n}},
  \bibinfo{author}{\bibfnamefont{S.}~\bibnamefont{Salomonson}},
  \bibnamefont{and} \bibinfo{author}{\bibfnamefont{I.}~\bibnamefont{Lindgren}},
  \bibinfo{journal}{Phys. Rev. A} \textbf{\bibinfo{volume}{65}},
  \bibinfo{pages}{032516} (\bibinfo{year}{2002}).

\bibitem[{\citenamefont{Edmonds}(1957)}]{Edm57}
\bibinfo{author}{\bibfnamefont{A.~R.} \bibnamefont{Edmonds}},
  \emph{\bibinfo{title}{\textit{Angular Momentum in Quantum Mechanics}}}
  (\bibinfo{publisher}{Princeton University Press},
  \bibinfo{address}{Princeton, N.J.}, \bibinfo{year}{1957}).

\bibitem[{\citenamefont{Lindgren and Ros\'{e}n}(1974)}]{LR74}
\bibinfo{author}{\bibfnamefont{I.}~\bibnamefont{Lindgren}} \bibnamefont{and}
  \bibinfo{author}{\bibfnamefont{A.}~\bibnamefont{Ros\'{e}n}},
  \bibinfo{journal}{Case Studies in Atomic Physics}
  \textbf{\bibinfo{volume}{4}}, \bibinfo{pages}{93} (\bibinfo{year}{1974}).

\bibitem[{\citenamefont{Arfken}(1985)}]{Arf84}
\bibinfo{author}{\bibfnamefont{G.}~\bibnamefont{Arfken}},
  \emph{\bibinfo{title}{\textit{Mathematical Methods for Physicists}}}
  (\bibinfo{publisher}{Academic Press}, \bibinfo{address}{San Diego},
  \bibinfo{year}{1985}).

\end{thebibliography}

 \end{document}